\documentclass[a4paper,11pt]{article}
\usepackage{jheppub}
\usepackage[T1]{fontenc}
\usepackage{amsmath,amssymb,amsfonts}
\usepackage{bm}
\usepackage{graphicx}
\usepackage{hyperref}
\usepackage{orcidlink}
\usepackage{enumitem}
\usepackage{physics}
\usepackage{caption}
\usepackage{tikz}
\usetikzlibrary{decorations.pathmorphing,decorations.markings,arrows.meta}
\usepackage{subcaption}
\usepackage{float}
\usepackage[colorinlistoftodos]{todonotes}
\arxivnumber{2605.27641}

\title{\boldmath Robin holography in AdS and BTZ: double-trace RG flow and exceptional points}          

\author[a]{Y. Wang \orcidlink{0009-0009-2952-5920}}
\author[a]{J. Yang \orcidlink{0009-0007-8425-4612}}
\affiliation[a]{Department of Physics, Brown University, Providence, RI, 02912, USA}

\emailAdd{yiru\_wang2@brown.edu}
\emailAdd{juanyi\_yang@brown.edu}

\abstract{We construct the exact Robin bulk-to-boundary propagator for a Breitenl\"ohner--Freedman scalar on AdS$_{d+1}$ and the BTZ black hole, realizing the double-trace RG flow between standard and alternate quantization geometrically as a one-parameter family of bulk boundary
conditions. We derive the UV and IR chain expansions of the kernel intrinsically from the boundary-value problem, without an auxiliary-field decoupling, and identify a branch split at each order that separates the local data the boundary CFT observes from finite-bulk-depth structure visible only to bulk probes -- the part of $K_f$ that distinguishes holographic reconstruction from boundary calculation.

On BTZ we obtain the closed-form Robin kernel and the corresponding family of quasinormal-mode trajectories, each connecting an alternate-quantization pole at $g=0$ to a standard one at $g\to\infty$. We locate an exceptional-point locus along this family at which two trajectories coalesce into a Jordan block, and show it acts as a non-Hermitian phase boundary for the double-trace flow itself: crossing it reorganizes the global pole-pairing topology of the spectrum. Unlike holographic EPs reached by analytic continuation in momentum or frequency, this transition lives on the interpolation between quantizations and is reachable at finite real momentum and temperature by tuning the physical Robin coupling.}

\begin{document}
\maketitle
\flushbottom

\section{Introduction}
\label{sec:intro}

A scalar field in $\text{AdS}_{d+1}$ with mass in the Breitenl\"ohner--Freedman
window $-d^2/4 \leq m^2 L^2 < -d^2/4 + 1$ admits two consistent quantizations
\cite{Breitenlohner:1982bm,Klebanov_1999}, with dual operators of dimension
$\Delta_\pm = d/2 \pm \nu$. The two are connected by a relevant double-trace
flow $\delta S = (f/2)\!\int\! \mathcal{O}_-^2$, running from the $\Delta_-$
fixed point in the UV to the $\Delta_+$ fixed point in the IR
\cite{witten2002multitraceoperatorsboundaryconditions,Berkooz_2002,Klebanov:2002ja}. From the
bulk side the flow is encoded entirely in the boundary condition: a Robin/mixed
condition $\gamma\,\phi + \epsilon\,\partial_z\phi = h$ at $z=\epsilon$
interpolates between $\gamma = -\Delta_-$ (alternate) and $\gamma\to\infty$
(standard), with intermediate $\gamma$ tuned as in \cite{Hartman_2008} so that
a finite double-trace coupling $f$ survives the $\epsilon\to 0$ limit. The
Robin family is therefore the realization of the double-trace flow in the bulk, and the bulk-to-boundary propagator $K_f$ at finite $f$ is a closed-form
scaling function that interpolates smoothly between the two fixed-point
kernels.

At finite temperature the flow has a thermal scale to compare against. For a
(1+1)d boundary, the BTZ black hole \cite{Banados:1992wn} is exactly solvable since
the radial equation is hypergeometric, the near-boundary coefficients
$\mathcal{A},\mathcal{B}$ are gamma function ratios, and the QNM condition reduces to
$\mathcal{B} + g\,\mathcal{A} = 0$ with $g$ a dimensionless Robin coupling
normalized at the thermal scale \cite{Son_2002}. Tracking the QNM spectrum as
$g$ runs from $0$ to $\infty$ visualizes the double-trace flow directly in the
complex frequency plane, with each trajectory connecting an alternate pole at
$g=0$ to a standard pole at $g\to\infty$.

This thermal flow carries a non-Hermitian feature that is not visible at zero
temperature. As $\nu$ is varied at fixed momentum, neighboring QNM
trajectories on the Robin family can collide at a critical coupling, producing
an exceptional point at which two damped frequencies coalesce into a Jordan
block \cite{Heiss_2012}. Crossing the EP locus reorganizes the global
pole-pairing topology, where the diagonal alternate-to-standard pairing seen at
small $\nu$ is replaced by a level-shifted one at large $\nu$. EPs have
appeared in holographic QNM spectra before
\cite{Arean:2019pom,Grozdanov:2019kge,Jansen_2020}, but as spectral features
within a fixed quantization, typically reached by analytic continuation in
$k$ or $\omega$. The Robin/BTZ EP is reachable by
tuning the physical double-trace coupling at finite real momentum, and its
locus has the meaning of an RG phase boundary.

The body of the paper develops the exact Robin holographic kernel and its UV/IR chain
expansions on pure AdS (Sections \ref{sec:robin}--\ref{sec:RGflow}) and then
specializes to BTZ for the QNM and EP analysis (Section \ref{sec:EP}). Two
points are worth noting. Firstly, the chain expansion is derived intrinsically from
the boundary-value problem rather than through a Hubbard--Stratonovich
decoupling \cite{Gubser_2003}, with the diagrammatic reading rests on large-$N$ factorization of
$\mathcal{O}^2$. The connection-formula split of of $K_f$ at each chain order into a
boundary-singular and a bulk-regular branch separates two physically distinct
pieces of holographic content meaningfully: the local OPE-like data that the boundary CFT
sees, and the finite-bulk-depth dressing visible only to probes that enter
the bulk — the part of $K_f$ that
distinguishes a genuine bulk reconstruction from a CFT calculation
rearranged in a particular way. Section \ref{sec:applications} closes with connections to
Luttinger-liquid crossover scaling, non-Hermitian holography, and coupled-SYK
constructions, and identifies the Lyapunov dependence on the Robin coupling as
an open direction.

\section{Robin Boundary Condition in AdS}
\label{sec:robin}
We consider a free scalar field $\phi$ of mass $m$ propagating in $(d{+}1)$-dimensional
conformally flat $AdS_{d+1}$ with metric

\begin{equation}\label{eq:metric}
  ds^2 = \frac{L^2}{z^2}\bigl(dz^2 + \eta_{\mu\nu}\,dx^\mu dx^\nu\bigr),
  \qquad z > 0\,.
\end{equation}
Near the conformal boundary $z\to 0$ the two linearly independent solutions of
the Klein--Gordon equation behave as
\begin{equation}\label{eq:asymp}
  \phi(x,z) \,\sim\, \beta(x)\,z^{\Delta_-} \,+\, \alpha(x)\,z^{\Delta_+}\,,
\end{equation}
with
\begin{equation}
  \Delta_\pm = d/2 \pm \nu\,,
  \qquad
  \nu = \sqrt{\frac{d^2}{4} + m^2 L^2}\,.
\end{equation}
The leading mode $z^{\Delta_-}$ is non-normalizable  and the subleading $z^{\Delta_+}$
is normalizable.

In the standard holographic dictionary one imposes boundary conditions at the regulated boundary $z = \epsilon$ and subsequently takes $\epsilon\to 0$ \cite{Skenderis:2002wp}.
The most general linear boundary condition that is first order in $z$-derivatives
is of the Robin/mixed type \cite{Compere:2008us,DelGrosso:2019gow}:
\begin{equation}\label{eq:robincondition}
  \gamma\,\phi(x,\epsilon) + \epsilon\,\partial_z\phi(x,\epsilon) = h(x)\,,
\end{equation}
where $\gamma$ is a real parameter and $h(x)$ is a prescribed boundary source.
This one-parameter family of boundary conditions flows bewteen standard/regular quantization
schemes and alternate/irregular one that are both allowed in the Breitenl\"ohner--Freedman window
$-d^2/4 \le m^2L^2 \le -d^2/4 + 1$ as limiting cases, and also interpolates between
them via a double-trace deformation of the dual CFT \cite{Breitenlohner:1982bm,witten2002multitraceoperatorsboundaryconditions,Berkooz_2002}.

Substituting the near-boundary expansion \eqref{eq:asymp} into \eqref{eq:robincondition}
gives
\begin{equation}\label{eq:robin_expanded}
  h(x) \,=\, (\gamma + \Delta_-)\,\beta(x)\,\epsilon^{\Delta_-}
         \,+\, (\gamma + \Delta_+)\,\alpha(x)\,\epsilon^{\Delta_+}
         \,+\, \cdots\,.
\end{equation}
The two standard quantization schemes are recovered as follows:

\begin{enumerate}[label=(\roman*)]
  \item Standard/regular quantization ($\gamma\to\infty$).
    The $\alpha$ term is suppressed by $\epsilon^{\Delta_+-\Delta_-}\to 0$,
    so the boundary condition fixes $\beta(x)$ as the source and leaves
    $\alpha(x)$ as the  response.
    The dual operator has conformal dimension $\Delta_+$.

  \item Alternate/irregular  quantization ($\gamma = -\Delta_-$).
    The coefficient of $\beta$ vanishes identically, so the boundary condition
    fixes $\alpha(x)$ as the source and $\beta(x)$ is the free response.
    The dual operator has conformal dimension $\Delta_-$.
\end{enumerate}

 For general $\gamma$, the source $h(x)$ is a linear combination of both $\alpha$ and $\beta$. This mixed boundary condtion can be precisely described by adding a double-trace deformation to the CFT action \cite{witten2002multitraceoperatorsboundaryconditions}
\begin{equation}
    S_{\text{CFT}} \to S_{\text{CFT}} + \frac{f}{2}\int d^dx\,\mathcal{O}(x)\,\mathcal{O}(x)\, ,
\end{equation}
where $f$ is a coupling related to $\gamma$  and this gives a modified equation of motion by new source
\begin{equation}
    J_{\mathrm{new}}(x) = \alpha(x)+ f \beta(x)\, .
\end{equation}
The theory is then off the conformal fixed point, flowing from the
$\Delta_-$ fixed point in the UV to the $\Delta_+$ fixed point in the IR \cite{witten2002multitraceoperatorsboundaryconditions,Berkooz_2002,Klebanov:2002ja}.

The previous holographic reconstruction of bulk from CFT dual \cite{Hamilton:2005ju,Hamilton:2006az} rely on being at a conformal fixed point where the boundary theory is a strict CFT with definite scaling dimension. 
However, the mixed condition deforms CFT by $f \mathcal{O}^2$, making it not conformal anymore. In general, it is not possible to have a well-defined conformal dimension since the boundary dual operator now becomes a mixture of $\mathcal{O}_{\Delta_-}$ and $\mathcal{O}_{\Delta_+}$.

Although given above constraints, we are still able to build the bulk solution from boundary source by simply formulating the problem as a boundary-value problem. This can be seen by the fact that regardless of the $\gamma$, the bulk reduces to the same radial ODE and what changes is only the UV boundary condition.

We now construct the bulk-to-boundary map associated with \eqref{eq:robincondition}. Going to Fourier space in the boundary directions, \begin{equation}
\phi(x,z) = \int \frac{\dd^d k}{(2\pi)^d}\,e^{ik\cdot x}\,\phi(k,z)\, ,
\end{equation}
the Klein--Gordon equation in the conformally flat coordinates becomes
\begin{equation}\label{eq:radialODE}
  \bigg(z^2\partial_z^2 - (d{-}1)\,z\,\partial_z - (k^2 z^2 + m^2 L^2)\bigg)
  \phi(k,z) = 0\,.
\end{equation}
Let $\psi(k,z)$ denote the solution of \eqref{eq:radialODE} that is
regular in the deep interior as $z\to\infty$, selecting
\begin{equation}\label{eq:psi}
  \psi(k,z) = z^{d/2}\,K_\nu(|k|z)\,,
\end{equation}
where $K_\nu$ is the modified Bessel function of the second kind.
(In a black-hole background, $\psi$ would instead be the ingoing solution at the horizon.)  The solution \eqref{eq:psi} is unique up to an overall $k$-dependent normalization.

The general on-shell bulk field compatible with interior regularity is
\begin{equation}
  \phi(k,z) = C(k)\,\psi(k,z)\,.
\end{equation}
Imposing the Robin condition \eqref{eq:robincondition} at $z = \epsilon$ determines
\begin{equation}
  C(k) = \frac{h(k)}{\gamma\,\psi(k,\epsilon) + \epsilon\,\psi'(k,\epsilon)}\,,
\end{equation}
where $\psi'(k,\epsilon) \equiv \partial_z\psi(k,z)\big|_{z=\epsilon}$.
Hence the full on-shell solution is
\begin{equation}\label{eq:phi_kernel}
  \phi(k,z) = K_\gamma(z,k)\,h(k)\,,
  \qquad
  K_\gamma(z,k) \,=\, \frac{\psi(k,z)}
      {\gamma\,\psi(k,\epsilon) + \epsilon\,\psi'(k,\epsilon)}\,.
\end{equation}
This is the holographic bulk-to-boundary kernel (or propagator) for
Robin parameter $\gamma$ at cutoff $\epsilon$.
In position space the map reads
\begin{equation}\label{eq:posspace}
  \phi(x,z) = \int \dd^d x'\,K_\gamma(z;\,x-x')\,h(x')\,,
\end{equation}
with $K_\gamma(z;\,x-x')$ the inverse Fourier transform of $K_\gamma(z,k)$.

\section{The deformed holographic map}
\label{sec:deform_btob}

We start from the Robin kernel
\begin{equation}
  K_\gamma(z,k) \,=\, \frac{\psi(k,z)}
      {\gamma\,\psi(k,\epsilon) + \epsilon\,\psi'(k,\epsilon)}\,,
\end{equation}
with $\psi(k,z) = z^{d/2} K_\nu(|k|z)$ the unique interior-regular solution. The small-argument expansion of $K_\nu$ gives
\begin{equation}
    \psi(k,\epsilon) =\epsilon^{d/2} K_{\nu}(|k|\epsilon) = c_-\,\epsilon^{\Delta_-}|k|^{-\nu}+c_+\,\epsilon^{\Delta_+}|k|^\nu+\cdots\,,
\end{equation}
with $c_- = 2^{\nu-1}\Gamma(\nu)$ and $c_+ = 2^{-\nu-1}\Gamma(-\nu)$. The boundary operator in the denominator then expands as
\begin{equation}
\label{eq:boundarysourceexpansion}
    \gamma\,\psi(k,\epsilon) + \epsilon\,\psi'(k,\epsilon) =(\gamma+\Delta_-)\,c_-\, \epsilon^{\Delta_-}|k|^{-\nu} +(\gamma+\Delta_+)\,c_+\,\epsilon^{\Delta_+}|k|^\nu+\cdots\,.
\end{equation}

The structure of \eqref{eq:boundarysourceexpansion} immediately distinguishes three regimes in $\gamma$, which we now treat in turn.

\subsection{Fine-tuned $\gamma$: balanced Robin parameter}
\label{subsec:balanced_robin_parameter}
For $\gamma+\Delta_-\neq 0$, the first term in \eqref{eq:boundarysourceexpansion} dominates as $\epsilon\to 0$, since $\Delta_-<\Delta_+$. Rescaling the source as $h(k) = \epsilon^{\Delta_-}\tilde h(k)$ to absorb the leading power, the renormalized kernel
\begin{equation}\label{eq:simplifiedkernel}
      \tilde{K}_{\gamma}(z,k) \,\equiv\, \epsilon^{-\Delta_-} K_\gamma(z,k)
      \,\xrightarrow{\epsilon\to 0}\,
      \frac{z^{d/2}K_\nu(|k|z)}{(\gamma+\Delta_-)\,c_-\, |k|^{-\nu}}
\end{equation}
collapses onto the undeformed standard ($\Delta_+$) bulk-to-boundary propagator: the $|k|^\nu$ branch is suppressed by $\epsilon^{2\nu}\to 0$ and drops out. In particular, the Dirichlet limit $\gamma\to\infty$ falls in this class.

A non-trivial mixing of the two branches in the $\epsilon\to 0$ limit requires the leading coefficient to be tuned to scale as $\epsilon^{2\nu}$, so that both terms in \eqref{eq:boundarysourceexpansion} contribute at the same order. Following \cite{Hartman_2008}, this is achieved by setting
\begin{equation}\label{eq:gammaf}
    \gamma=-\Delta_{-}-f\,\epsilon^{2 \nu}\left(2 \pi^{d / 2}\,\frac{\Gamma(1-\nu)}{\Gamma\left(\Delta_{-}\right)}\right)\,,
\end{equation}
where $f$ is the double-trace coupling. Substituting \eqref{eq:gammaf} into \eqref{eq:boundarysourceexpansion} cancels the leading $\epsilon^{\Delta_-}|k|^{-\nu}$ piece and replaces it with a finite $\epsilon^{\Delta_+}|k|^{-\nu}$ contribution at the same order as the $|k|^\nu$ branch. Rescaling the source as $h(k)=\epsilon^{\Delta_+}\tilde h(k)$ and stripping the common $\epsilon^{\Delta_+}$ factor, the renormalized Robin kernel reads
\begin{equation}\label{eq:Robin parameter f}
    \tilde{K}_{f}(z, k)=\frac{z^{d / 2} K_\nu(|k| z)}{-f\left(2 \pi^{d / 2}\frac{\Gamma(1-\nu)}{\Gamma\left(\Delta_{-}\right)}\right) c_{-}\,|k|^{-\nu}+2 \nu\, c_{+}\,|k|^\nu}\,.
\end{equation}
The denominator now contains both $|k|^\nu$ and $|k|^{-\nu}$ terms, with relative weight controlled by $f$. The two limiting cases of \eqref{eq:Robin parameter f} are: as $f\to 0$, the $|k|^\nu$ branch dominates and one recovers the alternate ($\Delta_-$) propagator; as $f\to\infty$, the $|k|^{-\nu}$ branch dominates and one recovers the standard ($\Delta_+$) propagator. The whole $f$-axis interpolates between the two CFT fixed points, in line with the double-trace RG flow interpretation of \cite{witten2002multitraceoperatorsboundaryconditions,Berkooz_2002}.

In what follows we suppress the overall $\epsilon$ factors and denote the renormalized kernel and source simply by $K_f$ and $h$ so that
\begin{equation}
    \label{eq:tuned_map}
    \phi(k,z) = K_f(z,k)\,h(k)\,.
\end{equation}

To make the structure of \eqref{eq:Robin parameter f} more transparent, define the scale
\begin{equation}\label{eq:muscale}
    \mu^{2\nu}\,\equiv\, -\,\frac{\left(2 \pi^{d / 2}\,\frac{\Gamma(1-\nu)}{\Gamma\left(\Delta_{-}\right)}\right)c_-}{2 \nu\, c_{+}}\,f\,,
\end{equation}
so that
\begin{equation}\label{eq:K_f}
    K_f(z, k)=\frac{z^{d / 2} K_\nu(|k| z)\,|k|^\nu}{2 \nu\, c_{+}\left(|k|^{2 \nu}+\mu^{2 \nu}\right)}\,.
\end{equation}
The dimensionful coupling is now packaged into a single scale $\mu$, with $[\mu]=\text{mass}$. We further introduce the regulated Euclidean invariant distance
\begin{equation}
    \label{eq:Euclideandistance}
    \bar{\xi} \,\equiv\, \xi z'\big|_{z'\to0}\,=\,\frac{z^2+r^2}{2z}\,,
\end{equation}
and its Lorentzian continuation
\begin{equation} \label{Lorentzdistance}
    \bar{\sigma} \,\equiv\, \sigma z'\big|_{z'\to0}\,=\,\frac{z^2-(t-t')^2+(\mathbf{x}-\mathbf{x'})^2}{2z}\,,
\end{equation}
restricted to spacelike separation, $\bar\sigma_+ \equiv \bar\sigma\,\Theta(\bar\sigma)$.

\subsection{UV expansion around $\Delta_-$}
\label{subsec:UV_expansion_around_Delta_-}
For $|k|>\mu$, the denominator of \eqref{eq:K_f} admits the convergent geometric expansion
\begin{equation}\label{eq:geomUV}
\frac{1}{|k|^{2\nu}+\mu^{2\nu}}=\sum_{n=0}^\infty
(-1)^n\,\mu^{2n\nu}\,|k|^{-2(n+1)\nu}\,,
\end{equation}
organizing $K_f$ as a power series as an expansion in the double-trace coupling around the UV ($\Delta_-$) fixed point. After Fourier transformation, the UV expansion can be written as
\begin{equation}
\label{eq:K_expansion_UV}
K_f(z,\bar\sigma_+)
=
\sum_{n=0}^\infty
(-f)^n K_f^{(n)}(z,\bar\sigma_+)
=
\sum_{n=0}^{\infty}
(-f)^n
C_n
\bigl[
A_n(z,\bar{\sigma}_+)+B_n(z,\bar{\sigma}_+)
\bigr] \,.
\end{equation}
Here the overall coefficient is
\begin{equation}
\label{eq:Cn}
C_n
=
\frac{2^{d/2-1-(2n+1)\nu}}
{2\nu c_+ (2\pi)^{d/2}}
\left(\frac{\mu^{2\nu}}{f}\right)^n \,.
\end{equation}

The hypergeometric connection formula splits each term in
\eqref{eq:K_expansion_UV} into two analytically distinct branches. The boundary-regular branch is
\begin{align}
\label{eq:UVAbranch}
A_n(z,\bar{\sigma}_+)
=
&\,z^{-d/2+(2n+1)\nu}
\frac{
\Gamma\left(d/2-n\nu\right)
\Gamma\left(d/2-(n+1)\nu\right)
\Gamma\left(-d/2+(2n+1)\nu\right)
}{
\Gamma(n\nu)\Gamma((n+1)\nu)
}
\notag\\
&\times
{}_2F_1\!\left(
d/2-n\nu,\,
d/2-(n+1)\nu;\,
1+d/2-(2n+1)\nu;\,
\frac{2\bar{\sigma}_+}{z}
\right)\,,
\end{align}
while the boundary-singular branch is
\begin{align}
\label{eq:UVBbranch}
B_n(z,\bar{\sigma}_+)
=
&\,(2\bar{\sigma}_+)^{-d/2+(2n+1)\nu}
\Gamma\left(d/2-(2n+1)\nu\right)
\notag\\
&\times
{}_2F_1\!\left(
n\nu,\,
(n+1)\nu;\,
1-d/2+(2n+1)\nu;\,
\frac{2\bar{\sigma}_+}{z}
\right)\,.
\end{align}
At $n=0$ the bulk-regular branch vanishes by  $1/\Gamma(0)=0$. The zeroth-order propagator is therefore captured entirely by $B_0$, while at $n\geq 1$ both branches contribute on equal footing.

In the boundary UV limit $|k|\to\infty$, every term in \eqref{eq:K_expansion_UV} beyond $n=0$ carries an additional factor of $(\mu/|k|)^{2\nu}\to 0$, so the series is dominated by the $n=0$ contribution. The propagator thus reduces to the undeformed conformal kernel 
\begin{equation}\label{eq:undeformedK-}
    K_{\Delta_-}(z,\bar\sigma_+)
    \propto
    (2\bar\sigma_+)^{-\Delta_-}
\end{equation}
of the UV fixed point, with operator dimension $\Delta_-$.

\subsection{IR expansion around $\Delta_+$}
\label{subsuc:IR_expansion_around_Delta_+}
For $|k|<\mu$, the denominator of \eqref{eq:K_f} admits the convergent geometric expansion
\begin{equation}\label{eq:geomIR}
\frac{1}{|k|^{2\nu}+\mu^{2\nu}} \,=\, \mu^{-2\nu}\sum_{n=0}^\infty (-1)^n\,\mu^{-2n\nu}\,|k|^{2n\nu}\,,
\end{equation}
which, given $\mu^{2\nu}\propto -f$, organizes $K_f$ as a power series in $(|k|/\mu)^{2\nu}$ -- equivalently, a series in $f^{-1}$ around the IR ($\Delta_+$) fixed point. After Fourier transformation we have
\begin{equation}\label{eq:K_expansion_IR}
   K_f(z, \bar{\sigma}_+)\,=\,\sum_{n=0}^{\infty}(-f)^{-(n+1)}\,\bar{C}_n\bigl[\bar{A}_n(z, \bar{\sigma}_+)+\bar{B}_n(z, \bar{\sigma}_+)\bigr]\,,
\end{equation}
where 
\begin{equation}
\label{eq:barCn}
\bar C_n
=
-
\frac{2^{(2n+1)\nu-1}}
{2\nu c_+\,\pi^{d/2}}
\left(
\frac{f}{\mu^{2\nu}}
\right)^{n+1}
\end{equation}
is the IR analogue of \eqref{eq:Cn}, obtained  by the same Fourier bookkeeping. The hypergeometric connection formula again splits each term into two branches:
\begin{align}\label{eq:IRAbranch}
 \bar A_n(z,\bar\sigma_+)
=
&\,
(2\bar\sigma_+)^{-d/2-(2n+1)\nu}
\Gamma\left(d/2+(2n+1)\nu\right)
\notag\\
&\times
{}_2F_1\!\left(
-n\nu,\,
-(n+1)\nu;\,
1-d/2-(2n+1)\nu;\,
\frac{2\bar\sigma_+}{z}
\right)\,,
\\
\label{eq:IRBbranch}
\bar B_n(z,\bar\sigma_+)
=
&\,
z^{-d/2-(2n+1)\nu}
\frac{
\Gamma\left(d/2+n\nu\right)
\Gamma\left(d/2+(n+1)\nu\right)
\Gamma\left(-d/2-(2n+1)\nu\right)
}{
\Gamma(-n\nu)\Gamma(-(n+1)\nu)
}
\notag\\
&\times
{}_2F_1\!\left(
d/2+n\nu,\,
d/2+(n+1)\nu;\,
1+d/2+(2n+1)\nu;\,
\frac{2\bar\sigma_+}{z}
\right)\,.
\end{align}
Compared with the UV branches, the role of $A_n$ and $B_n$ is mechanically reversed: in the IR expansion it is $\bar{A}_n$ that carries the boundary singularity $\bar\sigma_+^{-d/2-(2n+1)\nu}$, while $\bar B_n$ depends on the bulk depth $z$.

At $n=0$ the bulk-regular branch $\bar B_0$ vanishes for the same reason that the prefactor contains $1/\Gamma(0)=0$. The zeroth-order IR contribution is therefore captured entirely by $\bar A_0\propto \bar\sigma_+^{-d/2-\nu}=\bar\sigma_+^{-\Delta_+}$, which is precisely the undeformed standard ($\Delta_+$) bulk-to-boundary propagator
\begin{equation}\label{eq:undeformedK+}
    K_{\Delta_+}(z,\bar\sigma_+)
    \propto
    (2\bar\sigma_+)^{-\Delta_+}\,.
\end{equation}
This is consistent with the $|f|\to\infty$ (equivalently $\mu\to\infty$) limit of \eqref{eq:Robin parameter f}, where the $|k|^{-\nu}$ branch dominates and the kernel collapses onto the standard one.

The IR expansion is valid in the long-distance regime $|k|\ll\mu$, i.e.\ at large invariant separation $\bar\sigma_+\mu\gtrsim 1$. In this regime the $n=0$ term dominates because every higher order in \eqref{eq:K_expansion_IR} carries an extra factor of $(|k|/\mu)^{2\nu}\to 0$. The two-point function thus exhibits long-distance scaling controlled by $\Delta_+$, the dimension of the IR fixed point reached by the relevant double-trace flow.

\begin{figure}[h]
    \centering
    \includegraphics[width=\textwidth]{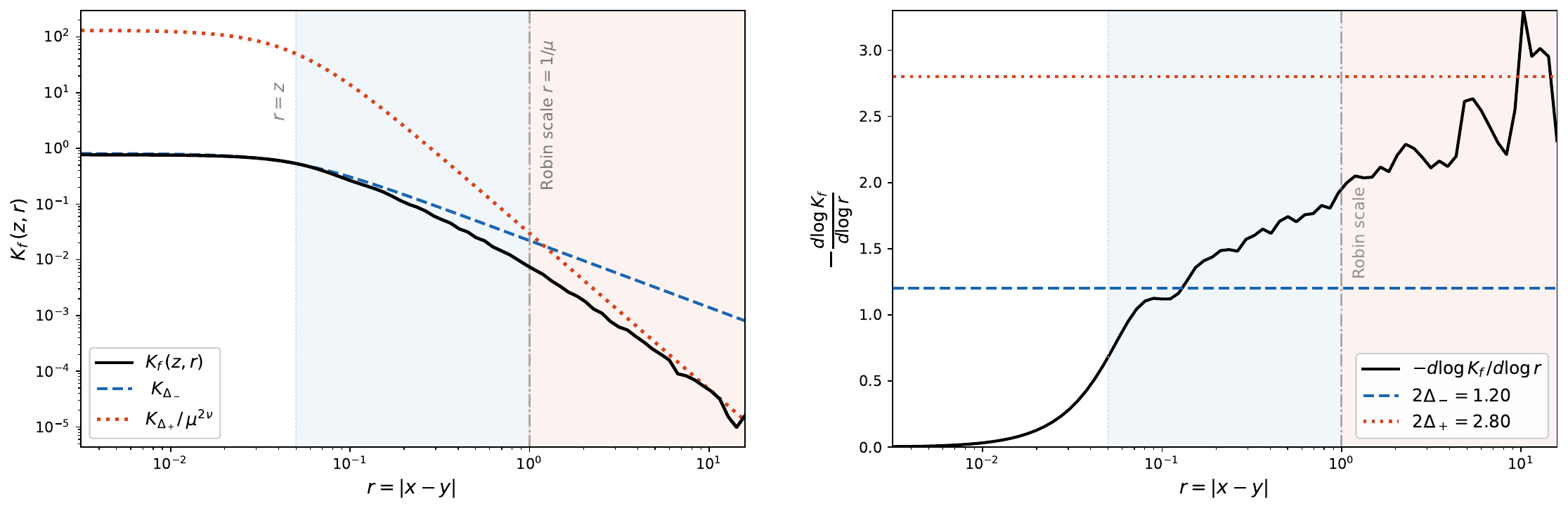}
    \caption{Position-space Robin kernel $K_f(z,r)$ in $d=2$ with $\nu=0.4$ ($\Delta_-=0.6$, $\Delta_+=1.4$) at bulk depth $z=0.05\ll 1/\mu$ and Robin scale $\mu=1$. Left: the exact kernel $K_f$ (solid) follows the undeformed alternate-quantization kernel $K_{\Delta_-}$ (dashed) at short distance and the standard-quantization kernel $K_{\Delta_+}/f$ (dotted) at long distance, joining smoothly across the Robin scale $r=1/\mu$ (dash-dotted vertical line). Right: the local power-law slope $-d\log K_f/d\log r$ runs continuously from $2\Delta_-=1.20$ in the UV regime to $2\Delta_+=2.80$ in the IR regime, with the crossover localized at $r\sim 1/\mu$.}
    \label{fig:Kf}
\end{figure}
\subsection{Crossover scale $k=|\mu|$}
\label{subsec:crossover_scale}
At this scale, neither series converges term by term at $|k|=\mu$. This  marks the boundary between the two regimes of validity, and is the natural crossover length $|x|\sim 1/\mu$ of the double-trace flow.

At the crossover, the exact kernel is nevertheless perfectly well defined,
\begin{equation}\label{eq:Kmu}
   K_f(z,\mu)   \,=\,\frac{z^{d/2}\,K_\nu(\mu z)}{4\nu\, c_+\,\mu^{\nu}}\,,
\end{equation}
and the crossover at $\bar\sigma_+\mu\sim 1$ is a smooth power-law transition between the two scaling regimes rather than an abrupt one: the effective slope $\partial\log|K_f|/\partial\log\bar\sigma_+$ runs continuously from $-\Delta_-$ at short distance to $-\Delta_+$ at long distance, with no exponential damping and no preferred intermediate exponent. This is illustrated in Fig. \ref{fig:Kf}, which evaluates the Hankel transform of \eqref{eq:K_f} numerically at fixed $\mu$ and small bulk depth $z\ll 1/\mu$. The left panel shows $K_f(z,r)$ alongside the two undeformed reference kernels $K_{\Delta_\pm}$, while the right panel extracts the local power-law slope $-d\log K_f/d\log r$, which interpolates monotonically and continuously from $2\Delta_-$ to $2\Delta_+$ across the Robin scale $r=1/\mu$.

\begin{figure}[h]
    \centering
    \includegraphics[width=0.6\textwidth] {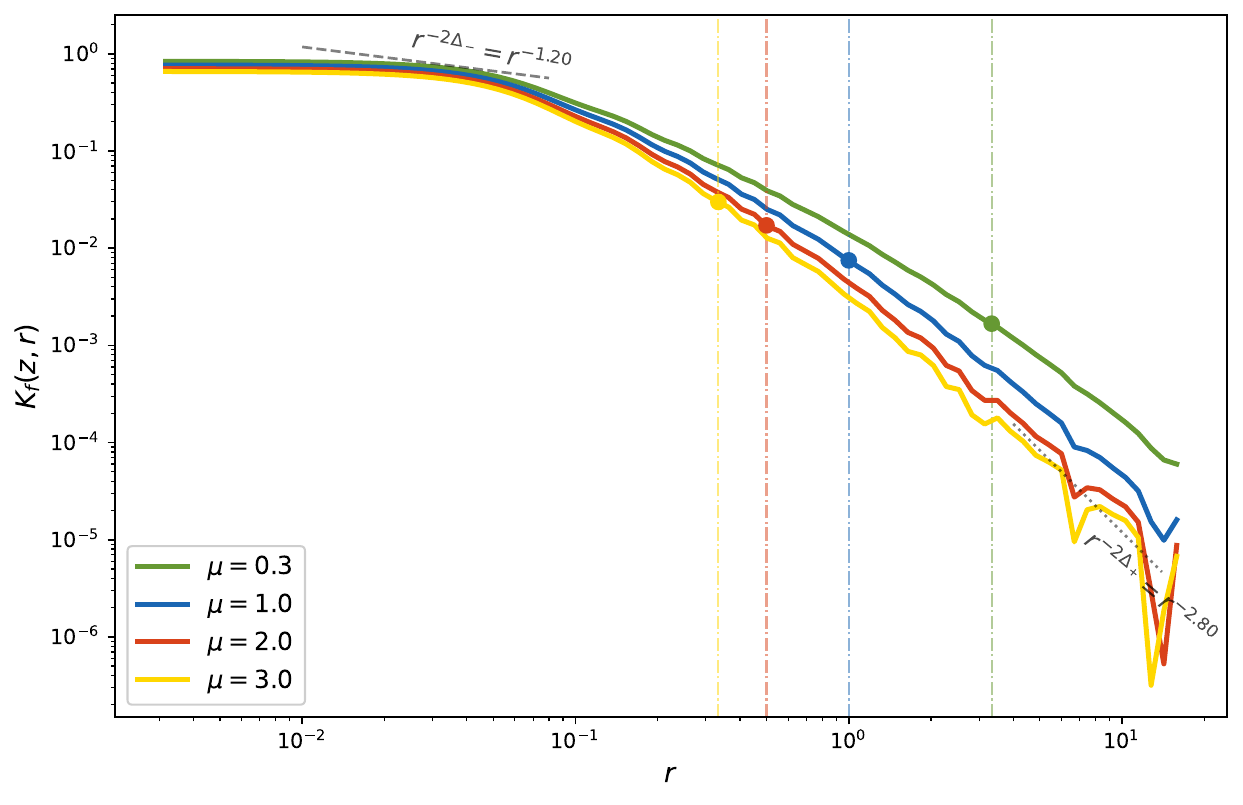}
    \caption{$\mu$-dependence of the crossover. For $\mu\in\{0.3,1,2,3\}$, each curve interpolates between the universal short-distance slope $r^{-2\Delta_-}=r^{-1.20}$ and long-distance slope $r^{-2\Delta_+}=r^{-2.80}$, with the turnover localized at the corresponding Robin scale $r=1/\mu$ (dots mark the crossover point on each curve). Tuning $\mu$ translates the crossover along the $r$-axis without altering the asymptotic dimensions, confirming that $\mu$ acts purely as an RG scale interpolating between the two CFT fixed points.}
    \label{fig:Kf_crossover}
\end{figure}

Varying $\mu$ moves the crossover but leaves the UV and IR slopes unchanged, as shown in Fig. \ref{fig:Kf_crossover}: each curve shares the same $r^{-2\Delta_-}$ behavior at short distance and the same $r^{-2\Delta_+}$ behavior at long distance, with the turnover localized at the respective Robin scale $r=1/\mu$. The crossover region is therefore precisely where the two expansions must be resummed into the exact mode-space kernel, and in this sense $|k|=\mu$ plays the role of a Robin RG scale separating the two CFT fixed points.
 
\section{Boundary physics from the deformed propagator}
\label{sec:boundarydata}

The position-space expansions \eqref{eq:K_expansion_UV} and
\eqref{eq:K_expansion_IR} encode the full spectral content of
the deformed boundary theory.  In this section we extract
the boundary OPE-like data order by order in $f$ directly from
the structure of $K_f$.

\subsection{Robin boundary two-point function}
\label{subsec:robin_boundary_2_point_function}

Let $q=\phi|_\epsilon$, $p=\epsilon\partial_z\phi|_\epsilon$, the bare action varies as $\delta S_{\text{on-shell}} = \int p\,\delta q$. The Legendre-type transform appropriate to fixing $h=p+\gamma q$ is
\begin{equation}
\label{eq:Robin-action}
S_\gamma = S_{\text{on-shell}} + \int \frac{d^dk}{(2\pi)^d}\left[-h(-k)q(k) + \tfrac{\gamma}{2}q(-k)q(k)\right],\end{equation}
such that
\begin{equation}
\delta S_\gamma = - \int \frac{d^dk}{(2\pi)^d}\,q(-k)\,\delta h(k).
\end{equation}
The one-point function conjugate to source $h$ is $\langle O(k)\rangle_h = -\alpha(k)$ after near-boundary rescaling in \eqref{eq:asymp}.

Since $\phi(z,k) = K_f(z,k)\,h(k)$, by reading off the $z^{\Delta_-}$ coefficient we obtain
\begin{equation}
\alpha(k) = \frac{c_-}{2\nu c_+\bigl(|k|^{2\nu}+\mu^{2\nu}\bigr)}\,h(k),
\end{equation}
and the formula above yields, up to a delta contact term,
\begin{equation}
\label{eq:deformedCFT2pt}
\langle O(k)O(-k)\rangle_f=\frac{\delta^2 S_\gamma}{\delta h(k) \delta h(-k)} =
\frac{2^{2\nu-1}\Gamma(\nu)}{\Gamma(1-\nu)}\,\frac{1}{|k|^{2\nu}+\mu^{2\nu}}.
\end{equation}
The Robin CFT two-point function we derive here shows the consistency with the one using the auxiliary field method in \cite{Gubser_2003}.

\subsection{Deformed CFT spectrum}
\label{subsec:deformed_CFT_spectrum}
The Robin CFT correlator
\begin{equation}
\label{eq:Robin-2pt}
G_f(k) \,=\, \frac{a_\nu}{|k|^{2\nu}+\mu^{2\nu}},
\qquad
a_\nu \equiv \frac{2^{2\nu-1}\Gamma(\nu)}{\Gamma(1-\nu)},
\end{equation}
interpolates between two scaling regimes set by the dynamical scale $\mu^{2\nu}\propto f$. Either regime admits a geometric expansion that resums into the full propagator and encodes the OPE-like data of a CFT perturbed by a relevant or irrelevant double-trace operator.

\paragraph{UV expansion ($|k|\gg\mu$).}
Expanding in $\mu^{2\nu}/|k|^{2\nu}$,
\begin{equation}
\label{eq:UV-momentum}
G_f(k) \,=\, a_\nu \sum_{n=0}^{\infty} (-1)^n\, \mu^{2n\nu}\, |k|^{-2\nu(n+1)}.
\end{equation}
Fourier transforming term by term using gives
\begin{equation}
\label{eq:UV-position}
\langle O(x)O(0)\rangle_f^{\rm UV}
\,=\, \frac{a_\nu}{\pi^{d/2}}\sum_{n=0}^{\infty} (-1)^n\, \frac{\Gamma(\Delta_- - n\nu)}{4^{\nu(n+1)}\,\Gamma(\nu(n+1))}\, \frac{\mu^{2n\nu}}{|x|^{2(\Delta_- - n\nu)}}.
\end{equation}
The $n=0$ term is the unperturbed two-point function of the alternate-quantization operator $O_-$ of dimension $\Delta_-$. The $n\geq 1$ terms are corrections in the dimensionless parameter $(\mu|x|)^{2\nu}$, organized as the spectral expansion of the relevant deformation the UV fixed point.

\paragraph{IR expansion ($|k|\ll\mu$).}
Expanding in $|k|^{2\nu}/\mu^{2\nu}$,
\begin{equation}
\label{eq:IR-momentum}
G_f(k) \,=\, \frac{a_\nu}{\mu^{2\nu}}\sum_{n=0}^{\infty} (-1)^n\, \mu^{-2n\nu}\,|k|^{2n\nu}.
\end{equation}
The $n=0$ piece is $k$-independent and contributes only a contact term. For $n\geq 1$, the same Fourier identity yields
\begin{equation}
\label{eq:IR-position}
\langle O(x)O(0)\rangle_f^{\rm IR}
\,=\, \frac{a_\nu}{\pi^{d/2}}\sum_{n=1}^{\infty}(-1)^n\, \frac{\Gamma(\Delta_+ + (n-1)\nu)}{4^{-n\nu}\,\Gamma(-n\nu)}\,\frac{1}{\mu^{2(n+1)\nu}\,|x|^{2(\Delta_+ + (n-1)\nu)}}.
\end{equation}
The leading $n=1$ term is the two-point function of the standard-quantization operator $O_+$ of dimension $\Delta_+$, suppressed by $f^{-2}$.  The $n\geq 1$ subleading terms are corrections in $1/(\mu |x|)^{2\nu}$, and represent irrelevant deformations of the standard-quantization IR fixed point controlled by the inverse double-trace coupling $1/f$.

\subsection{Chain diagrams}
\label{subsec:chainstructure}

The double-trace deformation $\delta S = f \int d^d x\,\mathcal{O}^2(x)$ can be implemented without modifying the boundary condition, by treating the undeformed alternate ($\Delta_-$) theory as the free theory and $f\mathcal{O}^2$ as a boundary interaction vertex \cite{witten2002multitraceoperatorsboundaryconditions, Sever_2002, muckwofgang2002}. The building ingredients are: the bulk-to-boundary propagator
\begin{equation}
\label{eq:undeformed btob propagtor}
K_{\Delta_-}(z,x;y)\propto\left(\frac{z}{z^2+|x-y|^2}\right)^{\Delta_-},
\end{equation}
connecting a bulk point $(z,x)$ to a boundary point $y$, and the boundary two-point function
\begin{equation}
\label{eq:Gdef}
G_{\Delta_-}(y_i,y_j)\propto\frac{1}{|y_i-y_j|^{2\Delta_-}},\qquad
\widetilde G_{\Delta_-}(k)\propto |k|^{2\Delta_- -d}=|k|^{-2\nu}.
\end{equation}

Each insertion of the double-trace vertex contributes one power of $f$ and one boundary integration $\int d^dy_i$. A key analytical advantage of the present formulation is that the geometric resummation of $K_f$ is derived intrinsically from the exact boundary-value problem, bypassing entirely the Hubbard–Stratonovich auxiliary-field trick used in earlier treatments \cite{Gubser_2003}. The resummation is exact at the level of the boundary-value problem; its interpretation as a sum of tree-level chain diagrams relies on large-$N$ factorization, since connected $n$-point functions of $\mathcal{O}$ scale as $\langle\mathcal{O}\cdots\mathcal{O}\rangle_{\rm c}\sim N^{2-n}$, so chain contractions at order $f^n$ are $O(1)$ while connected higher-order correlators are suppressed by $1/N^2$.

\subsubsection{UV regime: small-$f$ expansion}
\label{subsubsec:UV_regime}
For $|k|>\mu$ the geometric series
\begin{equation}
\label{eq:geomUV}
\frac{|k|^\nu}{|k|^{2\nu}+\mu^{2\nu}}
=\sum_{n=0}^\infty (-1)^n\,\mu^{2n\nu}\,|k|^{-(2n+1)\nu}
\end{equation}
converges. Using $\mu^{2\nu}\propto f$ and $\widetilde G_{\Delta_-}(k)=|k|^{-2\nu}$, the $n$-th term of the Robin kernel reads
\begin{equation}
\label{eq:nth_mode}
K_f^{(n)}(z,k)\propto (-f)^n\bigl(|k|^{-2\nu}\bigr)^n\times|k|^{-\nu}\,z^{d/2}K_\nu(|k|z).
\end{equation}
Each factor $|k|^{-2\nu}=\widetilde G_{\Delta_-}(k)$ is one boundary link in Fourier space, and the product of $n$ such factors becomes a boundary convolution chain in position space:
\begin{equation}
\label{eq:chaindiagram}
K_f^{(n)}(z,x;y)
=(-f_{\mathrm{eff}})^n \int d^dy_1\cdots d^dy_n\,
K_{\Delta_-}(z,x;y_1)\,G_{\Delta_-}(y_1,y_2)\cdots G_{\Delta_-}(y_n,y)\,.
\end{equation}

A chain of length $n$ has $n$ double-trace vertices $\{y_1,\dots,y_n\}$ along the boundary, with the bulk point connected to $y_1$ by $K_{\Delta_-}$ and the source point pinned at $y$ (see Fig. \ref{fig:chain_UV}). Summing over all chain lengths gives
\begin{equation}
\label{eq:Kgammachain}
K_f
=\sum_{n=0}^\infty K_f^{(n)}
=K_{\Delta_-}-f_{\mathrm{eff}}\,K_{\Delta_-} *G_{\Delta_-}+(f_{\mathrm{eff}})^2 K_{\Delta_-} *G_{\Delta_-} *G_{\Delta_-}+\cdots,
\end{equation}
with $*$ the boundary convolution.

\paragraph{Position-space branches of $K_f$.}

After Fourier transforming term by term, the hypergeometric connection
formula splits each $K_f^{(n)}$ into two independent structures.  One
branch, denoted $B_n$, is singular as the chordal distance
$\bar\sigma_+\to0$.  Its leading behavior is
\begin{equation}
\label{eq:Bn_leading}
B_n
\sim
\bar\sigma_+^{-d/2+(2n+1)\nu}
=
\bar\sigma_+^{-\Delta_n^{(B)}},
\qquad
\Delta_n^{(B)}
=
\frac d2-(2n+1)\nu
=
\Delta_- -2n\nu .
\end{equation}
Thus each additional chain link softens the boundary singularity by
$2\nu$.  The gamma-function coefficient $
\Gamma\left(\frac d2-(2n+1)\nu\right)$
encodes the strength of this singular contribution.  Since the effective
powers $\Delta_n^{(B)}=\Delta_- -2n\nu$ decrease with $n$, these terms are the OPE-like singular powers generated by the large-$N$ double-trace chain expansion.

The second branch, denoted $A_n$, is regular as
$\bar\sigma_+\to0$.  Its leading near-boundary-singularity behavior is
controlled by the bulk radial depth,
\begin{equation}
\label{eq:An_leading}
A_n
\sim
z^{-d/2+(2n+1)\nu}
\frac{
\Gamma(d/2-n\nu)\,
\Gamma(d/2-(n+1)\nu)\,
\Gamma(-d/2+(2n+1)\nu)
}{
\Gamma(n\nu)\,
\Gamma((n+1)\nu)
}.
\end{equation}
Because this branch remains finite when $\bar\sigma_+\to0$, it does not
contribute to the leading boundary short-distance singularity.  Instead, it
captures the part of the chain that is sensitive to finite bulk depth.  The
factor
$\left(
\Gamma(n\nu)\Gamma((n+1)\nu)
\right)^{-1}$
suppresses the extended-chain contribution at large $n$.  In particular,
the $A_n$ branch vanishes at $n=0$ due to the factor
$\Gamma(n\nu)^{-1}$, consistent with the fact that the zeroth-order term
contains no extended boundary chain.

\subsubsection{IR regime: large-$f$ expansion}
\label{subsubsec:IR_regime}

\begin{figure}[ht]
\centering

\begin{subfigure}[t]{\textwidth}
\centering
\begin{tikzpicture}[scale=1.0]
  \draw[thick,gray] (0,0) -- (10.2,0);
  \node[gray,font=\footnotesize,right] at (10.2,0) {$z=0$};

  \filldraw[black] (1,0) circle (2.5pt);
  \filldraw[black] (3,0) circle (2.5pt);
  \filldraw[black] (5,0) circle (2.5pt);
  \filldraw[black] (8,0) circle (2.5pt);

  \draw[thick,violet,decorate,decoration={snake,amplitude=1.5pt,segment length=6pt}]
    (1,0) -- (3,0);
  \draw[thick,violet,decorate,decoration={snake,amplitude=1.5pt,segment length=6pt}]
    (3,0) -- (5,0);
  \node[font=\small] at (6.2,0) {$\cdots$};
  \draw[thick,violet,decorate,decoration={snake,amplitude=1.5pt,segment length=6pt}]
    (7.2,0) -- (8,0);

  \draw[thick,blue] (8,0) -- (8,2.5);
  \filldraw[blue] (8,2.5) circle (3pt);

  \node[below,font=\footnotesize] at (1,-0.15) {$y$};
  \node[below,font=\footnotesize] at (3,-0.15) {$y_n$};
  \node[below,font=\footnotesize] at (5,-0.15) {$y_{n-1}$};
  \node[below,font=\footnotesize] at (8,-0.15) {$y_1$};

  \node[blue,font=\footnotesize,right] at (8.1,2.6) {$(z,\,x)$};
  \node[blue,font=\footnotesize,right] at (8.05,1.25) {$K_{\Delta_-}$};
  \node[violet,font=\footnotesize,above] at (2,0.15) {$G_{\Delta_-}$};

  \foreach \x in {1,3,5} {
    \node[font=\footnotesize] at (\x,0.45) {$-f$};
  }
\end{tikzpicture}
\caption{UV chain ($z < 1/\mu$): perturbation around the $\Delta_-$ fixed point}
\label{fig:chain_UV}
\end{subfigure}

\vspace{1.2em}

\begin{subfigure}[t]{\textwidth}
\centering
\begin{tikzpicture}[scale=1.0]
  \draw[thick,gray] (0,0) -- (10.2,0);
  \node[gray,font=\footnotesize,right] at (10.2,0) {$z=0$};

  \filldraw[black] (1,0) circle (2.5pt);
  \filldraw[black] (3,0) circle (2.5pt);
  \filldraw[black] (5,0) circle (2.5pt);
  \filldraw[black] (8,0) circle (2.5pt);

  \draw[thick,red,decorate,decoration={snake,amplitude=1.5pt,segment length=6pt}]
    (1,0) -- (3,0);
  \draw[thick,red,decorate,decoration={snake,amplitude=1.5pt,segment length=6pt}]
    (3,0) -- (5,0);
  \node[font=\small] at (6.2,0) {$\cdots$};
  \draw[thick,red,decorate,decoration={snake,amplitude=1.5pt,segment length=6pt}]
    (7.2,0) -- (8,0);

  \draw[thick,blue] (8,0) -- (8,2.5);
  \filldraw[blue] (8,2.5) circle (3pt);

  \node[below,font=\footnotesize] at (1,-0.15) {$y$};
  \node[below,font=\footnotesize] at (3,-0.15) {$y_n$};
  \node[below,font=\footnotesize] at (5,-0.15) {$y_{n-1}$};
  \node[below,font=\footnotesize] at (8,-0.15) {$y_1$};

  \node[blue,font=\footnotesize,right] at (8.1,2.6) {$(z,\,x)$};
  \node[blue,font=\footnotesize,right] at (8.05,1.25) {$K_{\Delta_+}$};
  \node[red,font=\footnotesize,above] at (2,0.15) {$G_{\Delta_+}$};

  \foreach \x in {1,3,5,7.5} {
    \node[font=\footnotesize] at (\x,0.45) {$-1/f$};
  }
\end{tikzpicture}
\caption{IR chain ($z > 1/\mu$): perturbation around the $\Delta_+$ fixed point}
\label{fig:chain_IR}
\end{subfigure}

\caption{\small Order-$n$ chain contribution to the Robin kernel
$K_f^{(n)}(z, x;\, y)$ in the two regimes. The diagrammatic
structure is identical in both panels: a bulk-to-boundary leg from the
bulk point $(z, x)$ down to $y_1$, and a chain of $n$ boundary
links running through $y_n, y_{n-1}, \ldots, y_1$ to the source $y$. The
two expansions differ only in which propagators dress the chain and in the
sign of the effective coupling: (a) for $z < 1/\mu$ the UV expansion uses
$K_{\Delta_-}$ and $G_{\Delta_-}$ with vertex factor $-f$, organizing the
chain as a perturbation around the $\Delta_-$ fixed point; (b) for
$z > 1/\mu$ the IR expansion uses $K_{\Delta_+}$ and $G_{\Delta_+}$ with
vertex factor $-1/f$, organizing the chain around the $\Delta_+$ fixed
point. The crossover at $z = 1/\mu$ is the bulk-depth image of the
boundary RG scale.}
\label{fig:chain_UV_IR}
\end{figure}
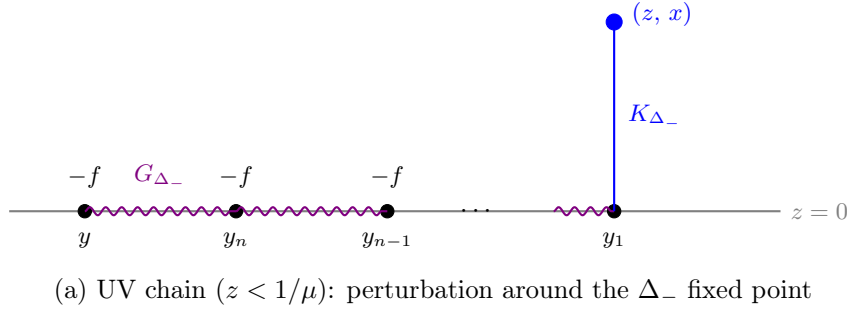
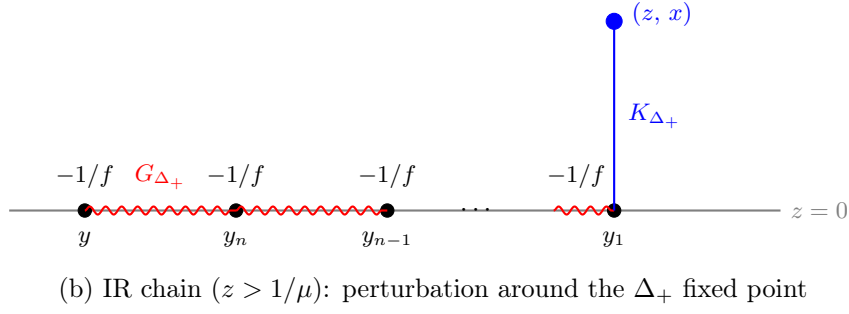

For $|k|<\mu$ the alternative geometric series
\begin{equation}
\label{eq:geomIR}
\frac{|k|^\nu}{|k|^{2\nu}+\mu^{2\nu}}
=\sum_{n=0}^\infty(-1)^n\mu^{-2(n+1)\nu}|k|^{(2n+1)\nu}
\end{equation}
converges. Each factor of $|k|^{2\nu}$ is now $\widetilde G_{\Delta_+}(k)$, the Fourier transform of the standard-quantization boundary propagator $G_{\Delta_+}(y_i,y_j)\propto|y_i-y_j|^{-2\Delta_+}$, and each $\mu^{-2\nu}\propto f^{-1}$ counts an insertion of the inverse coupling. The Robin kernel resums to a chain (see in Fig. \ref{fig:chain_IR}) around the IR fixed point,
\begin{equation}
\label{eq:KgammaIRchain}
K_f=\sum_{n=0}^\infty -(-f_{\mathrm{eff}})^{-(n+1)}\,K_{\Delta_+}*\underbrace{G_{\Delta_+} *\cdots*G_{\Delta_+}}_{n},
\end{equation}
with $1/f$ playing the role of vertex coupling. The connection formula again  separates the answer into a boundary-singular branch $\bar A_n$ and a bulk-regular branch $\bar B_n$ where:

$\bar A_n\sim\bar\sigma_+^{-(\Delta_+ +2n\nu)}$, recovering the standard-quantization propagator at $n=0$, with effective boundary dimension $\Delta_n^{(\bar A)}=\Delta_+ +2n\nu$;

$\bar B_n$ carries $\Gamma(-n\nu)^{-1}$, vanishing at $n=0$, and is regular at $\bar\sigma_+\to 0$ with leading $z$-dependence $z^{-d/2-(2n+1)\nu}$, encoding the finite bulk structure invisible to the boundary spectrum.

The bulk-regular branches $A_n$, $\bar B_n$ vanish at $n=0$ in both expansions, reflecting the absence of an extended chain at zeroth order; they encode finite-bulk-depth information invisible to the local boundary OPE-like expansion. Neither of them converges at the crossover $|k|=\mu$, where the dynamical scale equals the probe momentum; the exact kernel $K_f$ interpolates between the two regimes at the midpoint of the flow. As $f$ is dialed continuously from $0$ to $\infty$, the propagator passes smoothly between the $\Delta_-$ and $\Delta_+$ scaling regimes, with the crossover localized around $|k|\sim\mu(f)\propto f^{1/2\nu}$.

\subsubsection{Holographic interpretation of the chain diagrams}
\label{subsubsec:holographic_interpretation_of_the_chain_diagrams}

The chain diagrams admit a direct holographic reading once we observe that the bulk-to-boundary leg $K_{\Delta_\pm}(z, x; y)$ has boundary support of width $\sim z$ around the bulk projection $x$. The first chain integration is therefore confined to a boundary patch of size $z$, so bulk depth and boundary scale are matched at the level of the integrand, where a chain anchored at a bulk point of depth $z$ extends across exactly the boundary scale that depth $z$ would resolve. The holographic UV/IR connection is therefore built into the chain structure.

This identification determines which chain expansion is the relevant description at each bulk depth. For $z \ll 1/\mu$ the chain is confined within the UV slice $|y - x| < 1/\mu$, where the perturbative kernel is composed of $K_{\Delta_-}$ and the boundary link  $G_{\Delta_-}$, and the UV chain expansion perturbative in $f$ around the $\Delta_-$ fixed point converges. For $z \gg 1/\mu$ the chain is forced to extend past the Robin scale into the IR tail $|y - x| > 1/\mu$, where the kernel is built from $K_{\Delta_+}$ and $G_{\Delta_+}$, and the IR chain expansion perturbative in $1/f$ around the $\Delta_+$ fixed point takes over. The crossover $z \sim 1/\mu$ is the bulk-depth image of the boundary RG scale, with the choice between UV and IR descriptions on two side of this Robin scale.

Within whichever expansion applies, the connection-formula split $K_f^{(n)} = A_n (\bar{B}_n) + B_n (\bar{A}_n)$ separates each chain order into two physically distinct channels. $B_n (\bar{A}_n)$ is the branch that controls the near-coincidence limit; it carries the local boundary OPE-like singularity and feeds into $G_f$. $A_n (\bar{B}_n)$ comes from the extended-chain configuration, in which the internal vertices spread across the full scale-$z$ boundary patch; it depends on the bulk depth $z$ and contributes no boundary singularity. The non-trivial holographic content of the chain expansion sits in this second piece. The non-local boundary extension is the dual of the finite-bulk-depth structure it carries. The renormalized boundary two-point function discards them as contact terms, but they are precisely the part of $K_f$ that distinguishes a genuine bulk reconstruction from a purely CFT calculation rearranged in a particular way. The chain expansion thus provides a manifestly holographic decomposition of the deformed kernel where $B_n (\bar{A}_n)$ is what the boundary CFT directly observes, and $A_n (\bar{B}_n)$ is the dressing of the bulk interior the deformation has built up at depth $z$, accessible only to probes that go into the bulk.

\section{Holographic RG flow}
\label{sec:RGflow}

The one-parameter Robin family realizes the double-trace RG flow between the alternate ($\Delta_-$) and standard ($\Delta_+$) conformal fixed points geometrically, as a continuous deformation of the boundary condition at $z=\epsilon$ \cite{Papadimitriou_2007}. We now extract from the exact kernel \eqref{eq:K_f} an RG flow equation in the double-trace coupling $f$ \cite{Heemskerk:2010hk,Faulkner:2010jy} and show that the deformed boundary correlator drives the flow.

\subsection{Flow equation and recursive structure}
\label{subsec:flow_equation}

The exact mode-space kernel
\begin{equation}\label{eq:mukernel}
    K_f(z, k)=\frac{z^{d/2}\, K_\nu(|k| z)\,|k|^\nu}{2\nu\, c_{+}\bigl(|k|^{2\nu}+\mu^{2\nu}\bigr)}
\end{equation}
depends on $f$ only through $\mu^{2\nu}\propto f$. Rewriting $\mu^{2\nu} = \lambda f$
with $\lambda$ collecting all $f$-independent prefactors, a direct differentiation of \eqref{eq:mukernel} gives
\begin{equation}\label{eq:modespaceflowequation}
    \frac{\partial K_f(z,k)}{\partial f}\,=\,-\,\frac{\lambda}{a_\nu}\,\langle\mathcal{O}\mathcal{O}\rangle_f(k)\,K_f(z,k)\,.
\end{equation} 
The flow is multiplicative in mode space where each momentum mode evolves independently under $\partial_f$ with rate proportional to $\langle\mathcal{O}\mathcal{O}\rangle_f$ at that momentum. The boundary correlator thus same plays a double role as the observable on the boundary side and the rate that drives the bulk RG flows.

Fourier transforming \eqref{eq:modespaceflowequation} converts the mode-space product into a position-space convolution:
\begin{equation}\label{eq:Callan-Symanzik eqn}
    \frac{\partial K_f(z,\bar\sigma_+)}{\partial f}\,=\,-\,\frac{\lambda}{a_\nu}\!\int\! d^d x''\, \langle\mathcal{O}(x)\mathcal{O}(x'')\rangle_f\,K_f\bigl(z,\bar\sigma_+(x'',x')\bigr)\,,
\end{equation}
where $x$ is the boundary endpoint of $K_f$ on the left-hand side, $x'$ the bulk transverse position entering $\bar\sigma_+\equiv\bar\sigma_+(x,x')$, and $x''$ the convolution variable integrated over the boundary. This is the position-space Callan-Symanzik equation for the deformed bulk-to-boundary propagator where an infinitesimal change in $f$ smears $K_f$ along the boundary by the deformed two-point function.

Expanding both sides of \eqref{eq:Callan-Symanzik eqn} around the UV fixed point and matching powers of $f$ yields a strict recursion relation. At $f=0$ the deformed correlator reduces to the undeformed alternate-quantization two-point function $\langle\mathcal{O}\mathcal{O}\rangle_{\Delta_-}(k)\propto|k|^{-2\nu}$, and the leading matching gives
\begin{equation}\label{eq:recursiveintegral}
    (n+1)\,K_f^{(n+1)}(z,\bar\sigma_+)\,=\,-\,\frac{\lambda}{a_\nu}\!\int\! d^d x''\, \langle\mathcal{O}(x)\mathcal{O}(x'')\rangle_{\Delta_-}\,K_f^{(n)}\bigl(z,\bar\sigma_+(x'',x')\bigr)\,,
\end{equation}
with seed $K_f^{(0)}$ the undeformed alternate-quantization propagator. Each iteration adds one boundary integration weighted by the conformal two-point function $\langle\mathcal{O}\mathcal{O}\rangle_{\Delta_-}$ which gives the chain insertion of Section \ref{subsec:chainstructure}. Expanding instead around  $f=\infty$ produces the analogous recursion built on $\langle\mathcal{O}\mathcal{O}\rangle_{\Delta_+}$ for the IR chain.

\subsection{Boundary correlator}
\label{subsec:Boundary_correlator}

To make \eqref{eq:Callan-Symanzik eqn} more explicit in position space we Fourier transform the deformed boundary correlator. For general $\nu$ the result is expressible in Fox $H$-functions; at special values it reduces to elementary closed forms. For $\nu=1/2$,
\begin{equation}\label{eq:nu=1/2}
    \langle\mathcal{O}\mathcal{O}\rangle_f(x)\big|_{\nu=1/2}\,=\,\int\!\frac{d^d k}{(2\pi)^d}\,\frac{e^{ik\cdot x}}{|k|+\mu}\,\propto\,\frac{\mu}{(\mu^2 x^2+1)^{(d+1)/2}}\,,
\end{equation}
peaked at $x=0$ with width $\sim 1/\mu$ and a power-law tail. At the upper edge of the BF window, $\nu=1$,
\begin{equation}\label{eq:nu=1}
    \langle\mathcal{O}\mathcal{O}\rangle_f(x)\big|_{\nu=1}\,\propto\,\left(\frac{\mu}{|x|}\right)^{d/2-1}\!K_{d/2-1}(\mu|x|)\,,
\end{equation}
decaying exponentially as $e^{-\mu|x|}/|x|^{(d-1)/2}$ at large $|x|$. In both cases the boundary correlator is localized on the crossover scale $|x|\sim 1/\mu$, so the double-trace deformation modifies $K_f$ only at boundary separations comparable to the RG crossover length.

Using $\partial_f=(\partial_f\mu)\partial_\mu$, the flow equation takes the form
\begin{equation}\label{eq:muflow}
    \mu\,\frac{\partial K_f}{\partial\mu}\,=\,-\,\frac{2\nu\,\mu^{2\nu}}{a_\nu}\,\langle\mathcal{O}\mathcal{O}\rangle_f \,\star\, K_f\,,
\end{equation}
with $\star$ the position-space convolution. At the UV fixed point $\mu\to 0$ the right-hand side vanishes and $K_f$ becomes $\mu$-independent, consistent with pure $\Delta_-$ quantization. At the IR fixed point $\mu\to\infty$ the boundary correlator localizes,
\begin{equation}
    \langle\mathcal{O}\mathcal{O}\rangle_f(x)\,\xrightarrow{\mu\to\infty}\,\frac{a_\nu}{\mu^{2\nu}}\,\delta^{(d)}(x)\,,
\end{equation}
and \eqref{eq:muflow} collapses to an ordinary differential equation,
\begin{equation}\label{eq:IRfloweqn}
    \mu\,\frac{\partial K_f}{\partial\mu}\,\xrightarrow{\mu\to\infty}\,-\,2\nu\,K_f\,,
\end{equation}
whose solution $K_f\propto f^{-1}$ matches the leading IR scaling in \eqref{eq:K_expansion_IR}.

The same flow equation applied to the boundary correlator itself integrates rationally
\begin{equation}\label{eq:Callan-Symanzik boundary2pt}
    \frac{\partial}{\partial f}\langle\mathcal{O}\mathcal{O}\rangle_f(k)\,=\,-\,\frac{\lambda}{a_\nu}\,\langle\mathcal{O}\mathcal{O}\rangle_f(k)^2\,,\qquad\Longrightarrow\qquad \langle\mathcal{O}\mathcal{O}\rangle_f(k)\,\propto\,\frac{1}{|k|^{2\nu}+\lambda f}\,,
\end{equation}
reproducing \eqref{eq:deformedCFT2pt} with the integration constant fixed by the UV boundary condition. In particular, we derive the deformed correlator directly from the holographic flow, independently of the on-shell action  given in previous section.

\section{Robin/BTZ and exceptional points}
\label{sec:EP}

We work on the non-rotating BTZ black hole \cite{Banados:1992wn} in three dimensions,
\begin{equation}
\label{eq:BTZ-metric}
ds^2 = -f(r)\,dt^2 + \frac{dr^2}{f(r)} + r^2\,d\varphi^2,
\qquad
f(r) = \frac{r^2 - r_+^2}{\ell^2},
\end{equation}
where $\ell$ is the AdS$_3$ radius, $r_+$ is the horizon, and $\varphi\sim\varphi+2\pi$.
The Hawking temperature is $T = r_+/(2\pi\ell^2)$. We set $\ell = 1$ in what follows.
Wick rotation $t\to -i\tau$ with $\tau\sim\tau+1/T$ produces the Euclidean BTZ geometry, on which we analytically continue boundary correlators after solving in Lorentzian
signature with infalling boundary conditions at the horizon \cite{Son_2002,Skenderis:2008dg}. The resulting QNM spectrum coincides with the poles of the dual thermal CFT$_2$ correlator \cite{Birmingham_2002}.

A massive scalar $\phi$ of mass $m$ satisfies
\begin{equation}
\label{eq:KG}
\bigl(\Box - m^2\bigr)\phi = 0,
\qquad m^2 = \Delta_+\Delta_- = \nu^2 - 1,
\end{equation}
with $\Delta_\pm = 1\pm\nu$ the standard and alternate conformal dimensions of the dual
operator. We work throughout in the Breitenlöhner--Freedman window $0<\nu<1$, where both
quantizations are unitary and the Robin family interpolates between them.

Decomposing
\begin{equation}
\phi(t,r,\varphi) = \sum_{k\in\mathbb{Z}}\int \frac{d\omega}{2\pi}\,
e^{-i\omega t + ik\varphi}\,\psi_{\omega k}(r),
\end{equation}
the radial equation is conveniently written in the coordinate
$z = 1 - r_+^2/r^2 \in [0,1]$, which sends the horizon to $z=0$ and the conformal boundary
to $z=1$. In this coordinate $\psi$ satisfies
\begin{equation}
\label{eq:radial}
z(1-z)\,\psi'' + (1-z)\,\psi'
+ \biggl[\frac{\tilde{\omega}^2}{4\,z}
- \frac{\tilde{k}^2}{4\,(1-z)}
- \frac{1-\nu^2}{4(1-z)}\biggr]\psi = 0,
\end{equation}
where we have introduced the dimensionless variables
\begin{equation}
\tilde{\omega} \equiv \frac{\omega}{2\pi T},
\qquad
\tilde{k} \equiv \frac{k}{2\pi T}.
\end{equation}

Equation \eqref{eq:radial} is hypergeometric. The exponents at the horizon $z=0$ are
$\pm i\tilde{\omega}/2$, and the ingoing mode corresponds to $\psi \sim z^{-i\tilde{\omega}/2}$
as $z\to 0$. Writing
\begin{equation}
\psi(z) = z^{-i\tilde{\omega}/2}\,(1-z)^{(1-\nu)/2}\,F(z),
\end{equation}
turns \eqref{eq:radial} into the standard hypergeometric equation, with the unique
horizon-ingoing solution
\begin{equation}
\label{eq:exact-soln}
\psi^{\text{in}}(z) = z^{-i\tilde{\omega}/2}\,(1-z)^{(1-\nu)/2}\,
{}_2F_1 \left(\tfrac{1-\nu}{2}-\tfrac{i(\tilde{\omega}+\tilde{k})}{2},\,
                \tfrac{1-\nu}{2}-\tfrac{i(\tilde{\omega}-\tilde{k})}{2};\,
                1 - i\tilde{\omega};\, z\right).
\end{equation}
This is the BTZ analogue of the standard infalling Eddington-Finkelstein solution; by
construction it is purely ingoing at the horizon and is therefore the appropriate
analytic continuation of the Euclidean regular solution to the lower-half-$\omega$
plane that defines the retarded correlator.

Near the boundary, the horizon-ingoing solution decomposes into the two
boundary falloffs as
\begin{equation}
\label{eq:near-bdy}
\psi^{\text{in}}(z) \,=\, A(\omega,k;\nu)\,(1-z)^{\Delta_-/2}
\,+\, B(\omega,k;\nu)\,(1-z)^{\Delta_+/2} \,+\, \cdots,
\end{equation}
with the omitted terms subleading within each branch. The coefficients
follow from the ${}_2F_1$ connection formula; absorbing the universal
horizon factor $\Gamma(1-i\tilde\omega)$ and the connection-formula
$\Gamma(\pm\nu)$ into a common normalization through
$A = \Gamma(1-i\tilde\omega)\Gamma(\nu)\,\mathcal{A}$ and
$B = \Gamma(1-i\tilde\omega)\Gamma(-\nu)\,\mathcal{B}$, the rescaled
coefficients are
\begin{equation}
\label{eq:AB}
\mathcal{A}(\omega,k;\nu)
=\frac{1}{\Gamma\!\left(\tfrac{1+\nu}{2}-\tfrac{i(\tilde\omega+\tilde k)}{2}\right)
        \Gamma\!\left(\tfrac{1+\nu}{2}-\tfrac{i(\tilde\omega-\tilde k)}{2}\right)},
\qquad
\mathcal{B}(\omega,k;\nu)
=\frac{1}{\Gamma\!\left(\tfrac{1-\nu}{2}-\tfrac{i(\tilde\omega+\tilde k)}{2}\right)
        \Gamma\!\left(\tfrac{1-\nu}{2}-\tfrac{i(\tilde\omega-\tilde k)}{2}\right)},
\end{equation}
both entire in $\omega$.

The two terms in \eqref{eq:near-bdy} are the alternate and standard
falloffs at dimensions $\Delta_-$ and $\Delta_+$. The Robin condition
mixes them at the dimensionless coupling
$g(f) = \Gamma(\nu)^2\sin(\pi\nu)\,f$, producing the master equation
\begin{equation}
\label{eq:master_eq}
H(\tilde\omega;\,g,\,\tilde k,\,\nu)
\,\equiv\,
\mathcal{B}(\tilde\omega,\tilde k;\nu)
\,+\, g\,\mathcal{A}(\tilde\omega,\tilde k;\nu) \,=\, 0,
\end{equation}
whose zeros locate the QNMs along the Robin family. This is the bulk boundary-condition realization of the double-trace spectral flow, whose pole motion was studied in the spectral-duality framework \cite{Grozdanov:2025ulc}.

The same condition with a non-zero source determines the
bulk-to-boundary kernel. Writing
$\phi(z;\omega,k) = C(\omega,k)\,\psi^{\text{in}}(z;\omega,k)$ and applying \eqref{eq:robincondition} with the same $\epsilon$-rescaling that
kept $f$ finite in Section  \ref{sec:deform_btob} gives
$C(\omega,k)\bigl[\mathcal{B} + g\,\mathcal{A}\bigr] = h(\omega,k)$,
where $h \equiv \tilde h /[\,2\nu\,\Gamma(1-i\tilde\omega)\Gamma(-\nu)\,]$
is the renormalized source absorbing the universal prefactors. Solving for $C$ we obtain
\begin{equation}
\label{eq:Robin-BTZ-kernel}
K_g^{\text{BTZ}}(z;\,\omega,k)
\,=\,
\frac{\psi^{\text{in}}(z;\omega,k)}
     {\mathcal{B}(\omega,k;\nu) + g\,\mathcal{A}(\omega,k;\nu)},
\end{equation}
the closed-form BTZ analogue of \eqref{eq:K_f}, with poles in $\omega$ at
the Robin QNMs of Fig. \ref{fig:qnm-trajectories} by construction. The
position-space kernel
\begin{equation}
\label{eq:Robin-BTZ-position}
K_g^{\text{BTZ}}(z;\,t,\varphi)
\,=\,
\sum_{k\in\mathbb{Z}} \int\!\frac{d\omega}{2\pi}\,
e^{-i\omega t + ik\varphi}\, K_g^{\text{BTZ}}(z;\omega,k)
\end{equation}
has no elementary closed form at finite $g$, but the angular
periodicity $\varphi\sim\varphi+2\pi$ of BTZ reorganizes it as the image
sum
\begin{equation}
\label{eq:Robin-BTZ-imagesum}
K_g^{\text{BTZ}}(z;\,t,\varphi)
\,=\,
\sum_{n\in\mathbb{Z}}\,
K_g^{(\mathrm{cov})}(z;\,t,\,\varphi+2\pi n)
\end{equation}
over the unwrapped covering space, with $K_g^{(\mathrm{cov})}$ the
Robin propagator on the cover. The thermal data sit in the sum, and the
Robin deformation is in the $g$-dependence of each image.

\subsection{QNM trajectories in the complex frequency plane}
\label{subsec:EP_setup}
\begin{figure}[htbp]
    \centering
     \includegraphics[width=0.8\textwidth]{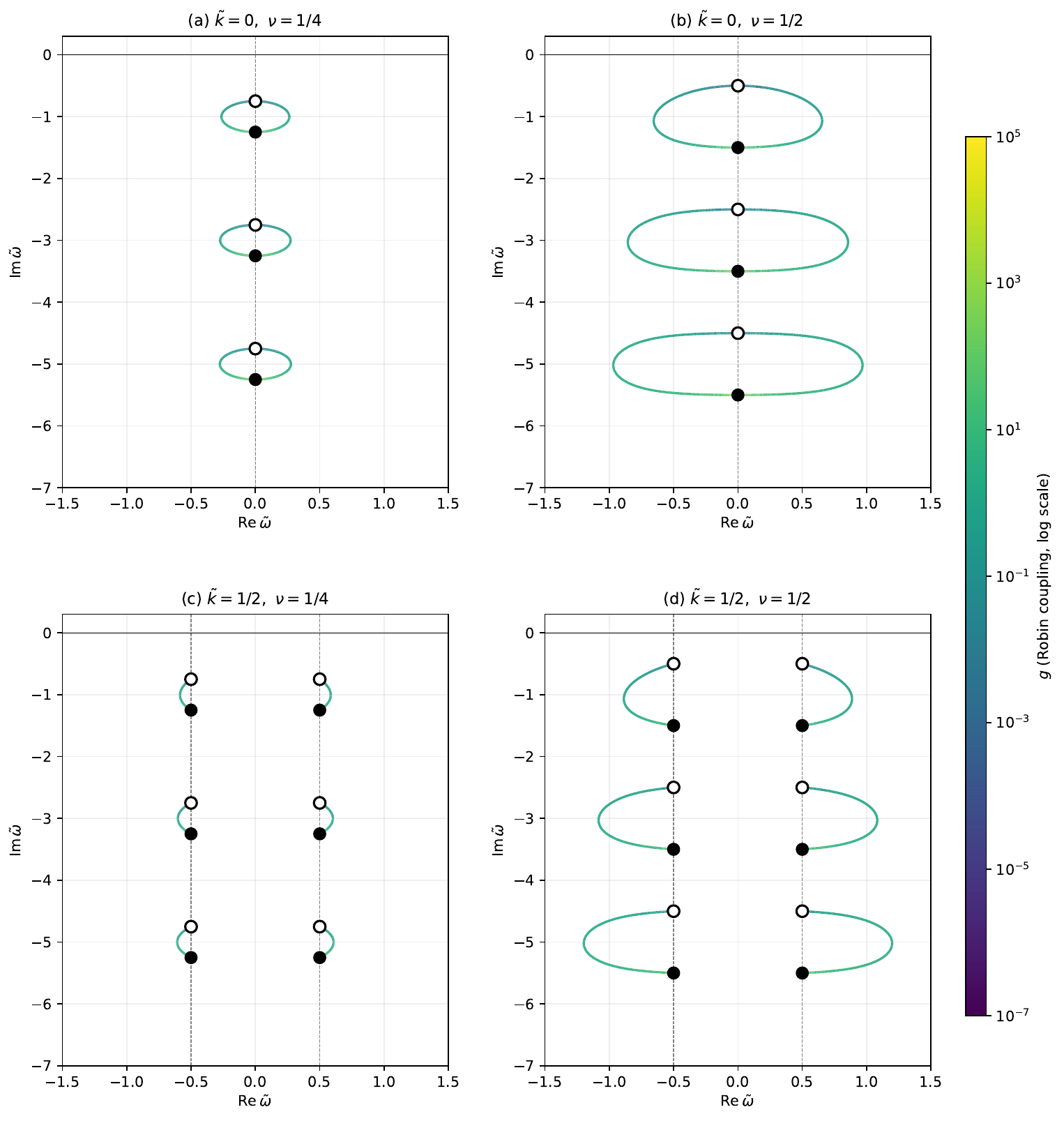}
    \caption{
    QNM trajectories in the complex $\tilde\omega$-plane as the Robin coupling
    $g$ is varied. Open circles denote the alternate-quantization poles at
    $g=0$; filled circles denote the standard-quantization poles reached as
    $g\to\infty$. The four panels compare
    $(\tilde k,\nu)=(0,1/4),(0,1/2),(1/2,1/4),(1/2,1/2)$.
    For $\tilde k=0$ each trajectory traces a closed loop centered on the
    imaginary axis; the loops arise because both endpoint poles are double
    zeros and the Robin perturbation splits each into a $\pm\sqrt{g}$ pair.
    For $\tilde k\neq 0$ the left- and right-moving towers split onto the chirality lines $\mathrm{Re}\,\tilde\omega = \pm \tilde k$, and the trajectories are open arcs connecting each alternate pole to its same-level standard partner. Larger $\nu$ produces wider excursions in real frequency, since the imaginary-axis separation $\Delta_+-\Delta_-=2\nu$ between alternate and standard towers grows.
    }
    \label{fig:qnm-trajectories}
\end{figure}
It is useful to regard the solutions as trajectories
\begin{equation}
\tilde{\omega}_j=\tilde{\omega}_j(g;\tilde{k},\nu)
\end{equation}
in the complex $\tilde{\omega}$-plane.  At $g=0$ the poles sit at the alternate-quantization spectrum
\begin{equation}
    \tilde{\omega}^{(\mathrm{alt})}_{m,\pm}
    =
    \pm \tilde{k}
    - i(1-\nu+2m),
    \qquad m=0,1,2,\ldots ,
\end{equation}
whereas for $g\to\infty$ they approach the standard-quantization spectrum \cite{Cardoso:2001hn}
\begin{equation}
    \tilde{\omega}^{(\mathrm{st})}_{n,\pm}
    =
    \pm \tilde{k}
    - i(1+\nu+2n),
    \qquad n=0,1,2,\ldots .
\end{equation}
Thus each Robin trajectory connects an alternate pole, drawn as an open circle
in Fig. \ref{fig:qnm-trajectories}, to a standard pole, drawn as a filled
circle.  The color of the curve records the value of $g$ along the flow.

Two qualitative features of Fig. \ref{fig:qnm-trajectories} deserve emphasis.
At $\tilde k = 0$ (panels (a), (b)), both gamma factors in $\mathcal B$ and
$\mathcal A$ coincide, so the endpoint QNMs are double zeros. The
Robin perturbation lifts the degeneracy via a square-root splitting
$\tilde\omega - \tilde\omega_0 \sim \pm\sqrt{g}$, and the two emerging branches
together trace a closed loop through both half-planes before re-merging at
the standard double zero as $g\to\infty$. These loops are the spectral
signature of boundary EPs sitting at the endpoints of the Robin family. For $\tilde k\neq 0$ (panels (c), (d)) the endpoints are simple
zeros, the loops open into smooth arcs, and the alternate-to-standard pairing
becomes diagonal: $\mathcal O^{R/L}_{\Delta_-,m}$ flows to $\mathcal O^{R/L}_{\Delta_+,m}$ within each chirality and at the same level.
This is the BTZ-thermal image of the standard Klebanov--Witten interpolation
between alternate and standard quantization of a BF-window scalar
\cite{Klebanov_1999}.

Increasing $\nu$ at fixed $\tilde k$, the arcs in panels (c), (d) bend outward and can collide with arcs at neighboring trajectory. At the collision two
QNM branches become equal and the local map $g \mapsto \tilde\omega(g)$ fails to be analytically separable. This is an exceptional point in QNM spectra\cite{Arean:2019pom,Heiss_2012}, and the next subsection traces what happens to the global trajectory structure across it.

\subsection{EP transition and pole-pairing rearrangement}
\label{subsec:EP_transition}
\begin{figure}[t]
    \centering
    \includegraphics[width=\textwidth]{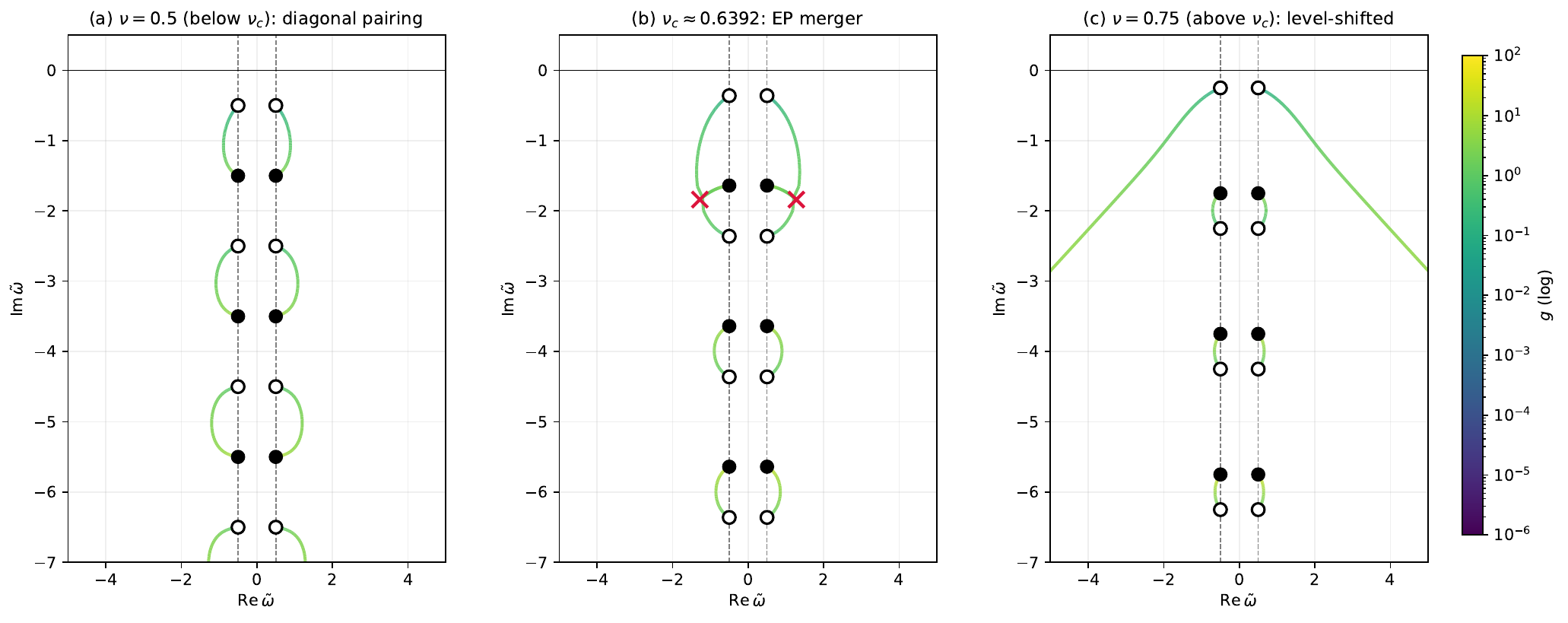}
    \caption{
    Exceptional-point transition at $\tilde k=1/2$.
    Panel (a): diagonal pairing below the critical value $\nu_c$. Panel (b):
    coalescence at the EP, marked by the red crosses at
    $\tilde\omega_c\simeq\pm 1.277-1.839\,i$ for $g_c\simeq 1.287$ and
    $\nu_c\simeq 0.6392$. Panel (c): level-shifted pairing above $\nu_c$;
    the visible short arcs pair the $m\geq 1$ rungs across the EP-induced
    shift, while the lightest ($m=0$) alternate pole exits the figure along
    a long arc and reconnects to a standard partner outside the displayed
    region.
    }
    \label{fig:EP-transition}
\end{figure}

Figure \ref{fig:EP-transition} shows the trajectory pattern at $\tilde k=1/2$ on three sides of the EP. Below a critical value $\nu_c$ the Robin flow preserves the diagonal pairing of Fig. \ref{fig:qnm-trajectories}(d). At $\nu = \nu_c \simeq 0.6392$, two adjacent arcs on each chirality line merge at order-2 exceptional points marked by the red crosses. Above $\nu_c$ the trajectories have been rerouted by the EP crossing: the $m\geq 1$ alternate
poles connect to level-shifted standard partners, while the $m=0$ alternate pole on each chirality is orphaned and connects to its standard endpoint only via a long arc that exits the figure and rejoins at much larger $|\mathrm{Re}\,\tilde\omega|$.

Near the collision, the master equation satisfies the simultaneous conditions
\begin{equation}
\label{eq:EP-conditions}
    H(\tilde\omega_c;g_c,\tilde k,\nu_c)=0,
    \qquad
    \partial_{\tilde\omega}H(\tilde\omega_c;g_c,\tilde k,\nu_c)=0,
\end{equation}
so that the two local solutions obey
\begin{equation}
\label{eq:sqrt-branch}
    \tilde\omega_\pm(g)
    = \tilde\omega_c
    \pm \mathcal C\sqrt{g-g_c}
    + O(g-g_c),
    \qquad
    \mathcal C =
    \left[-\frac{2\,\partial_g H}
    {\partial_{\tilde\omega}^2 H}\right]_{\mathrm{EP}}^{1/2}.
\end{equation}
The square-root structure is the local reason why the global trajectory
pattern reorganizes across the EP. Encircling the EP once in the
complexified $g$-plane exchanges the two branches; only after two encirclements
does each branch return to itself. The EP is therefore not a transverse
intersection of two curves but a branch point of the QNM spectrum, with
nontrivial monodromy in $g$ .

The physical content of the two sides is as follows. Below $\nu_c$, the
Robin flow connects poles in a way that preserves the original near-neighbor
pairing between the alternate and standard towers. At $\nu_c$, two damped
modes coalesce into a defective spectral point. Above $\nu_c$, the
monodromy around the EP has rerouted the global trajectories so that the
pole which began near one alternate-quantization level ends at a different
standard-quantization level. This is the finite-temperature BTZ realization
of a non-Hermitian spectral rearrangement along the double-trace RG flow.

For real Robin coupling, the local structure \eqref{eq:sqrt-branch} undergoes
a $\pi/2$ phase rotation across $g = g_c$: the square root $\sqrt{g - g_c}$
acquires a factor of $i$ as $g$ passes through the critical value, so the
displacement $\tilde\omega_\pm - \tilde\omega_c$ rotates by a right angle in
the complex plane. In the
language of non-Hermitian spectral theory the EP at $g = g_c$ acts as a
phase boundary on the real-$g$ line: the spectral order parameter
$(\tilde\omega_+ - \tilde\omega_-)^2 = 4\mathcal C^2(g - g_c)$ changes sign
there, and the two phases on either side are distinguished by the
orientation of the QNM pair in the complex frequency plane.

\subsection{EP locus}
\label{subsec:EPlocus}

\begin{figure}[htbp]
    \centering
     \includegraphics[width=0.9\textwidth]{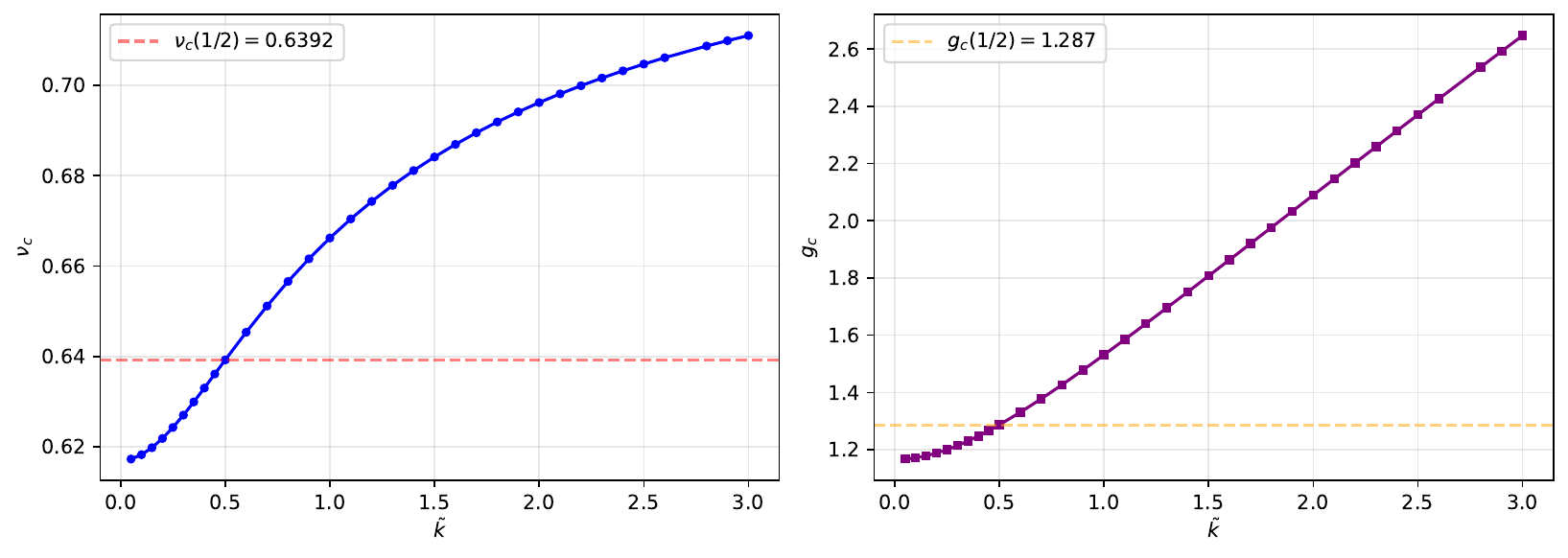}
    \caption{
    Exceptional-point locus as a function of dimensionless momentum
    $\tilde k$. Left: $\nu_c(\tilde k)$, ranging from
    $\nu_c\simeq 0.617$ at $\tilde k\to 0^+$ to $\nu_c\simeq 0.711$ at
    $\tilde k = 3$. Right: $g_c(\tilde k)$, rising monotonically from
    $g_c\simeq 1.16$ to $g_c\simeq 2.65$ over the same range and approaching
    approximately linear growth at large $\tilde k$. The benchmark point
    $\tilde k=1/2$, $\nu_c\simeq 0.6392$, $g_c\simeq 1.287$ is marked by
    dashed horizontal lines on both panels.
    }
    \label{fig:EP-locus}
\end{figure}

Imposing the EP conditions \eqref{eq:EP-conditions} on a grid of momenta and
solving by four-dimensional Newton continuation in
$(\mathrm{Re}\,\tilde\omega,\mathrm{Im}\,\tilde\omega,g,\nu)$ from the
$\tilde k=1/2$ benchmark traces out the EP locus shown in
Fig. \ref{fig:EP-locus}.

Several features are worth emphasizing. First, $\nu_c(\tilde k)$ is smooth and
monotonic in the plotted range. Larger
$\nu$ increases the imaginary-axis spacing between the standard and alternate
towers, and the EP occurs only once this spacing is large enough for
neighboring Robin trajectories to collide. Second, $g_c(\tilde k)$ also
increases with momentum, approximately linearly for $\tilde k \gtrsim 1$.
This reflects the fact that momentum separates the chirality lines
$\mathrm{Re}\,\tilde\omega=\pm\tilde k$, so a stronger Robin coupling is
required to bend trajectories from these lines into the region of the complex
plane where they can collide. Third, the $\tilde k\to 0$ limit of the locus
connects continuously to the boundary EPs visible in
Fig. \ref{fig:qnm-trajectories}(a, b): as $\tilde k\to 0$ the finite-$g_c$
EP migrates inward and merges with the endpoint degeneracies at $g=0$ and
$g=\infty$, which is consistent with the closed-loop trajectories observed
at $\tilde k = 0$.

The locus $(\tilde k,\nu_c(\tilde k),g_c(\tilde k))$ therefore acts as a phase boundary in the space of spectral flows. On one side, the Robin
deformation smoothly interpolates between alternate and standard quantization without changing the topology of the pole connections. On the other side,
the interpolation passes through a square-root branch point that reorganizes the pairing of QNM levels. In this sense the EP locus is the finite-temperature BTZ analogue of an exceptional-point phase diagram for the double-trace flow \cite{witten2002multitraceoperatorsboundaryconditions,Berkooz_2002}.

\subsection{Local EP physics at the benchmark point}
\label{subsec:EP_benchmark}

We now focus on the representative point
\begin{equation}
    \tilde k=\tfrac12,
    \qquad
    \nu=\nu_c\simeq 0.6392,
    \qquad
    g=g_c\simeq 1.287,
\end{equation}
where the right-moving pair coalesces at
$\tilde\omega_c\simeq 1.277-1.839\,i$. The left-moving partner sits at
$-\mathrm{Re}\,\tilde\omega_c + i\,\mathrm{Im}\,\tilde\omega_c$ by the
$\tilde k \to -\tilde k$ reflection symmetry. At the right-moving EP the
two QNM frequencies coalesce, and so do their radial wavefunctions: the
radial problem at $g=g_c$ no longer has two independent eigenfunctions at
the same frequency, but instead develops a Jordan block of length two.

After projecting onto the two colliding QNM branches, the effective
non-Hermitian Hamiltonian takes the universal $2\times 2$ form \cite{Heiss_2012}
\begin{equation}
    H_{\mathrm{eff}}^{(2)}(g)
    = \tilde\omega_c\,\mathbf 1
    + \begin{pmatrix} 0 & \beta \\ \alpha(g-g_c) & 0 \end{pmatrix}
    + O \bigl((g-g_c)^2\bigr),
\end{equation}
with eigenvalues
\begin{equation}
    \tilde\omega_\pm(g)
    = \tilde\omega_c \pm \sqrt{\alpha\beta\,(g-g_c)}
    + O(g-g_c).
\end{equation}
The off-diagonal $\beta$ is a normalization-dependent constant that can be
absorbed by a similarity transformation; the spectrally invariant content
is the product $\alpha\beta$, which is fixed by the second derivatives of
the exact Robin master equation at the EP,
\begin{equation}
    \alpha\beta
    = -\frac{2\,\partial_g H}{\partial_{\tilde\omega}^2 H}
    \bigg|_{\tilde\omega=\tilde\omega_c,\,g=g_c},
\end{equation}
and is the same coefficient $\mathcal C^2$ appearing in
Eq. \eqref{eq:sqrt-branch}. The global trajectory plots and the local
Taylor expansion therefore describe the same object from complementary
viewpoints: the plots show the large-scale rerouting of QNM branches, while
the local expansion shows that the rerouting is controlled by a square-root
branch point with universal $2\times 2$ Jordan structure.

The EP has a direct time-domain interpretation through the retarded
correlator
\begin{equation}
    G_R(t) \supset
    \sum_n c_n\,e^{-i\tilde\omega_n t}.
\end{equation}
Away from the EP, this sum contains two nearby damped oscillatory
contributions $c_\pm e^{-i\tilde\omega_\pm t}$. At the EP the two QNM
poles coalesce into a single second-order pole, and the inverse Fourier
transform produces a Jordan-block prefactor
\begin{equation}
    G_R(t)\big|_{\mathrm{EP}}
    \supset
    (c_0+c_1 t)\,e^{-i\tilde\omega_c t}.
\end{equation}
Since $\mathrm{Im}\,\tilde\omega_c\simeq -1.84 < 0$, the exponential damping
dominates the linear-in-$t$ growth and the late-time response decays with a characteristic single time scale. This is the
observable imprint of exceptional-point physics in the Robin BTZ quasinormal spectrum: two nearby damped relaxation channels of the retarded CFT$_2$ correlator merge into a single defective QNM channel where the resulting late-time response carries the characteristic Jordan-block prefactor.

\section{Applications and outlook}
\label{sec:applications}

The Robin/BTZ propagator developed in this paper is a closed-form holographic
realization of a (1+1)d CFT at finite temperature deformed by a relevant or
irrelevant double-trace operator, together with the explicit QNM spectrum of
that deformed theory. We close by sketching three settings in which this
construction is potentially useful, and one open question it leaves
unresolved.

\subsection{Crossover scaling in Luttinger liquids and quantum Hall edges}
\label{subsec:quantumhalledges}

The most direct phenomenological analogue of the Robin/BTZ crossover is
tunneling through a single impurity in a Luttinger liquid, or equivalently
through a quantum point contact between fractional quantum Hall edge
modes \cite{PhysRevB.46.15233,PhysRevB.41.12838}. The low-energy sector is a (1+1)d CFT at
temperature $T$ deformed by a tunneling operator that is relevant
($\Delta_-$) for one sign of the Luttinger parameter and irrelevant
($\Delta_+$) for the other, so that the RG flow connects two fixed points
with dimensions $\Delta_\pm$. Conductance interpolates between
$G\propto T^{2\Delta_--2}$ at weak tunneling and $G\propto T^{2\Delta_+-2}$
at strong tunneling \cite{PhysRevB.46.15233}, with the Luttinger parameter playing
the role of $\nu$. Exact results are available only on the perturbative
sides and at the free-fermion point $g=1/2$ \cite{Fendley1995}, while the
crossover scale, probed experimentally by point contacts in fractional
quantum Hall edges \cite{RevModPhys.75.1449} and carbon-nanotube
junctions \cite{Bockrath_1999}, is precisely where perturbative RG breaks
down on both sides.

The Robin/BTZ propagator is a large-$N$ holographic construction rather
than a microscopic description of any specific wire, but it does provide
an explicit scaling function -- the gamma-function ratio
$\mathcal{A}(\tilde\omega,\tilde k;\nu)/\mathcal{B}(\tilde\omega,\tilde k;\nu)$
of \eqref{eq:AB} -- that smoothly interpolates between the two fixed points
with no additional fitting parameters beyond $\nu$ and the dimensionless
coupling $\lambda = f\,T^{-2\nu}$. This makes it a natural benchmark for
the crossover shape against which numerical and experimental data can be
compared.

\subsection{Exceptional points and non-Hermitian holography}
\label{subsec:non-Hermitianholography}

The EP transition of Section \ref{sec:EP} suggests a qualitative spectral
feature that may have counterparts in finite-$N$ systems with similar
double-trace flow structure. At the EP, two damped QNMs coalesce at
$\tilde\omega_c\simeq 1.277 - 1.839\,i$ (for the benchmark
$\tilde k = 1/2$), and the retarded correlator at $g=g_c$ acquires a
Jordan-block prefactor $(c_0 + c_1 t)\,e^{-i\tilde\omega_c t}$. Across the
EP at fixed $\nu>\nu_c$, the global QNM trajectory pattern reorganizes as
the system passes through this branch point. Such non-Hermitian spectral
coalescences are studied broadly in open quantum and classical
systems \cite{Heiss_2012,Alu2019} and have also appeared in holographic
QNM spectra in different
settings \cite{Arean:2019pom,Grozdanov:2019kge,Jansen_2020}.

A point worth emphasizing is that the EP we identify lies on a
codimension-2 locus in the real $(\tilde k,\nu,g)$ parameter space, and
all three parameters are physical: $\tilde k$ is set by the boundary
kinematics, $\nu$ by the bulk mass, and $g$ by the Robin/double-trace
coupling. The EP is therefore reachable by tuning physical parameters at
finite temperature, not only as a feature of analytically continued
spectral curves. If an analogous spectral coalescence occurs in a
microscopic (1+1)d system with two competing tunneling channels, the
holographic calculation gives a concrete setting in which its structure
can be studied analytically. This might inform the interpretation of
finite-$N$ models where non-Hermitian spectral features are harder to
track.

\subsection{Connection to SYK, traversable wormholes, and chaos}
\label{subsec:connectiontoSYK}

A related setting in which the Robin/BTZ propagator may be useful is the
two-coupled-SYK model, which at large $N$ is dual to a nearly-AdS$_2$
geometry \cite{Maldacena_2016,maldacena2018eternaltraversablewormhole} with a double-trace
deformation $f\,\mathcal{O}_L\,\mathcal{O}_R$ between the two boundaries.
The Maldacena--Qi construction \cite{maldacena2018eternaltraversablewormhole}
identifies this deformation as the mechanism rendering the eternal
wormhole traversable, in the same sense that the Gao--Jafferis--Wall
protocol renders an ordinary two-sided black hole traversable via a
non-local boundary coupling \cite{Gao_2017}.

While the (1+1)d Robin/BTZ setup of this paper differs in dimension from
the (0+1)d SYK construction, the underlying analytic mechanism is the
same: a bulk-to-boundary kernel whose Robin parameter $f$ encodes
the strength of a relevant double-trace coupling between two CFT sectors.
The QNM spectrum and the EP structure derived here are direct analogues
of the spectral data one would compute on the AdS$_3$ side of an
analogous coupled-CFT$_2$ system.

The chaotic dynamics of such coupled systems is diagnosed by the
out-of-time-order correlator (OTOC),
\begin{equation}
\label{eq:OTOC4pt}
    C(t) \,=\, -\langle [W(t),V(0)]^2 \rangle_\beta
    \,\sim\, \frac{1}{N}\,e^{\lambda_L t},
\end{equation}
whose late-time growth rate $\lambda_L$ is bounded above by the
Maldacena--Shenker--Stanford bound $2\pi T$ \cite{Maldacena_2016}. In
the holographic eikonal computation of the OTOC, 
$K_f^{\text{BTZ}}$ appears as the kernel attaching external operators to the bulk shockwave, so the Robin deformation enters through the external-leg wavefunctions. The eikonal phase is a separate
shockwave-on-shockwave amplitude so  the QNM data of
Section \ref{sec:EP}  enter $\lambda_L(\lambda)$ through the
external legs but do not by themselves determine it.

How the Lyapunov exponent depends on the Robin coupling $\lambda$ is left open. Two limiting checks are clear: at $\lambda\to\infty$ the Robin family reaches its standard-quantization endpoint and one recovers the undeformed BTZ result saturating the MSS bound, $\lambda_L = 2\pi T$; at $\lambda\to 0$ the family reaches its alternate-quantization endpoint and the eikonal computation runs on the corresponding $\Delta_-$ background. The interpolation between these two limits, and in particular the behavior of $\lambda_L$  near the EP locus, would require an eikonal resummation in the Robin/BTZ background, which we leave for future work. The QNM spectrum derived in this paper supplies the input data such a computation needs.

\section{Summary}
\label{sec:summary}
The Robin boundary condition packages the double-trace RG flow geometrically.
The exact bulk-to-boundary kernel $K_f$ in the BF window depends on the coupling only through the single scale $\mu^{2\nu}\propto f$, and its
position-space behavior crosses over between $\Delta_-$ and $\Delta_+$ scaling
at $|x|\sim 1/\mu$. The two convergent geometric expansions of the kernel, valid for $|k|>\mu$ and $|k|<\mu$, resum as chain diagrams around the UV and IR fixed points; the derivation runs through the exact boundary-value problem, and its reading as a sum of tree-level chains rests on large-$N$ factorization of $\mathcal{O}^2$. At each chain order the connection formula separates the
answer into a boundary-singular branch carrying the local OPE-like data the CFT directly observes, and a bulk-regular branch carrying finite-bulk-depth structure that is absent from the renormalized two-point function but is the
part of $K_f$ a genuine bulk reconstruction sees. The same flow is captured at the level of $K_f$ by a position-space Callan--Symanzik equation in which the deformed two-point function plays the dual role of observable and rate, and
iterating it around either fixed point reproduces the chain recursion.

On BTZ, the Robin coupling generates a one-parameter family of QNM trajectories that, for generic $\nu$ and $\tilde k$, pair each alternate pole with the standard pole of the same chirality and level. Increasing $\nu$ at fixed $\tilde k$ brings two adjacent trajectories into collision at a critical coupling, and a Jordan block forms. The EP locus $(\nu_c(\tilde k),g_c(\tilde k))$, traced here by four-dimensional Newton continuation along $\tilde k\in(0,3]$, is a phase boundary for the global pole-pairing topology. Across EPs, the diagonal pairing is replaced by a
level-shifted one. On the locus the retarded correlator carries the universal $(c_0 + c_1 t)\,e^{-i\tilde\omega_c t}$ Jordan-block prefactor, with the coefficient fixed by second derivatives of the master equation.

What the Robin/BTZ propagator delivers, in closed form, is a (1+1)d thermal
CFT deformed by a relevant or irrelevant double-trace operator, its full QNM
spectrum along the flow, and a non-Hermitian spectral transition reachable by
tuning physical parameters at finite real temperature and momentum.

\appendix
\section{$\Delta_-$ expansion of $K_f$}
\label{sec:UV correction}

This appendix derives the position-space  expansion of $K_f$ around $\Delta_-$ point in detail, starting from the momentum-space kernel \eqref{eq:K_f}.

In momentum space, the deformation enters through the boundary factor
\begin{equation}
    \frac{|k|^\nu}{|k|^{2\nu}+\mu^{2\nu}} .
\end{equation}
In the UV regime $|k|^{2\nu}>\mu^{2\nu}$, this factor admits the convergent
geometric expansion
\begin{equation}
\label{eq:UVgeometricseriesofk}
    \frac{|k|^\nu}{|k|^{2 \nu}+\mu^{2 \nu}}
    \,=\,
    |k|^{-\nu}\,\frac{1}{1+\mu^{2\nu}|k|^{-2\nu}}
    \,=\,
    \sum_{n=0}^{\infty}
    (-1)^n\,\mu^{2 n \nu}\,|k|^{-(2 n+1) \nu},
\end{equation}
organized in powers of $(\mu/|k|)^{2\nu}$.

Substituting into \eqref{eq:K_f} and Fourier-transforming term by term using the standard identity \cite{GradshteynRyzhik} for radial integrals of
$K_\nu(|k|z)|k|^{\alpha}$, one obtains
\begin{align}
\label{eq:explicitK_gamma}
    K_f(z, r)
    \,=\,
    &\frac{1}{2 \nu c_{+}} 
    \sum_{n=0}^{\infty}\,
    \mathcal{N}_n\,(-\mu^{2\nu})^n\,
    \frac{
    \Gamma\!\left(d/2-n\nu\right)
    \Gamma\!\left(d/2-(n+1)\nu\right)
    }{
    \Gamma(d/2)
    }\,
    z^{-d/2+(2n+1)\nu}\notag\\
    &\qquad\qquad\times\,
    {}_2F_1\!\left(
    d/2-n\nu,\;
    d/2-(n+1)\nu;\;
    d/2;\;
    -\frac{r^2}{z^2}
    \right),
\end{align}
with the $n$-dependent prefactor
\begin{equation}\label{eq:Nn-def}
    \mathcal{N}_n
    \,\equiv\,
    \frac{2^{d/2-1-(2n+1)\nu}}{(2\pi)^{d/2}}\,.
\end{equation}
Here we have written
\begin{equation}
    r^2 \,=\, |\mathbf{x}-\mathbf{x}'|^2-(t-t')^2,
\end{equation}
in Lorentzian signature with smearing support restricted to spacelike
separation; in Euclidean signature one simply has $r^2=|\mathbf{x}-\mathbf{x}'|^2$.
The factor $\Theta(\bar\sigma_+)$ is suppressed throughout to lighten notation.

For $n=0$, the hypergeometric function simplifies via
${}_2F_1(a,b;a;x)=(1-x)^{-b}$, giving
\begin{equation}
    {}_2F_1\!\left(d/2,\,d/2-\nu;\,d/2;\,-\frac{r^2}{z^2}\right)
    \,=\,
    \left(1+\frac{r^2}{z^2}\right)^{-(d/2-\nu)}
    \,=\,
    \left(1+\frac{r^2}{z^2}\right)^{-\Delta_-}.
\end{equation}
Using $\Delta_\pm=d/2\pm\nu$ and $2\bar\sigma_+ = (z^2+r^2)/z$, the leading
term becomes
\begin{equation}\label{eq:Kf0final}
    K_f^{(0)}(z,r)
    \,=\,
    \frac{1}{2 \nu c_{+}}\,
    \frac{2^{\Delta_- -1}\,\Gamma(\Delta_-)}{(2\pi)^{d/2}}\,
    (2\bar\sigma_+)^{-\Delta_-}\,,
\end{equation}
which is proportional to the undeformed alternate-quantization
bulk-to-boundary propagator $K_{\Delta_-}$. The higher-order terms are
suppressed by $(\mu/|k|)^{2n\nu}$ in momentum space and encode the
deformation away from the UV fixed point.

To make the higher-order terms transparent in position space, apply the
${}_2F_1$ connection formula \cite{DLMF}
\begin{align}
{}_2F_1(a, b; c; z) 
\,=\,
&\frac{\Gamma(c)\Gamma(c-a-b)}{\Gamma(c-a)\Gamma(c-b)}\,
{}_2F_1(a, b; a+b-c+1; 1-z)\notag\\
&+\,
(1-z)^{c-a-b}\,
\frac{\Gamma(c)\Gamma(a+b-c)}{\Gamma(a)\Gamma(b)}\,
{}_2F_1(c-a, c-b; c-a-b+1; 1-z),
\end{align}
with
\begin{equation}
    a=d/2-n\nu,\qquad b=d/2-(n+1)\nu,\qquad c=d/2.
\end{equation}
The hypergeometric argument transforms as
\begin{equation}
    1-\left(-\frac{r^2}{z^2}\right)
    \,=\,
    \frac{z^2+r^2}{z^2}
    \,=\,
    \frac{2\bar\sigma_+}{z}\,.
\end{equation}
After collecting factors, the kernel splits into two branches:
\begin{equation}\label{eq:KfUVsigmabranches}
K_f(z,\bar{\sigma}_+)
\,=\,
\frac{1}{2\nu c_+}\sum_{n=0}^{\infty}
(-\mu^{2\nu})^n\,\mathcal{N}_n\,\bigl[A_n(z,\bar{\sigma}_+) + B_n(z,\bar{\sigma}_+)\bigr],
\end{equation}
where
\begin{align}
\label{eq:An-app}
A_n(z,\bar\sigma_+)
\,=\,
&\,z^{-d/2+(2n+1)\nu}\,
\frac{\Gamma\!\left(d/2-n\nu\right)\Gamma\!\left(d/2-(n+1)\nu\right)\Gamma\!\left(-d/2+(2n+1)\nu\right)}{\Gamma(n\nu)\,\Gamma((n+1)\nu)}\notag\\
&\times\,{}_2F_1\!\left(d/2-n\nu,\,d/2-(n+1)\nu;\,1+d/2-(2n+1)\nu;\,\frac{2\bar\sigma_+}{z}\right),\\
\label{eq:Bn-app}
B_n(z,\bar\sigma_+)
\,=\,
&\,(2\bar\sigma_+)^{-d/2+(2n+1)\nu}\,\Gamma\!\left(d/2-(2n+1)\nu\right)\notag\\
&\times\,{}_2F_1\!\left(n\nu,\,(n+1)\nu;\,1-d/2+(2n+1)\nu;\,\frac{2\bar\sigma_+}{z}\right).
\end{align}

The decomposition \eqref{eq:KfUVsigmabranches} separates each order of the UV
expansion into a boundary-singular piece ($B_n$, organized in powers of
$\bar\sigma_+$) and a bulk-regular piece ($A_n$, carrying the explicit
$z$-dependence). At $n=0$, $A_0$ vanishes by a gamma-function where
the prefactor contains $1/[\Gamma(0)\Gamma(\nu)]=0$. The leading UV
contribution is therefore captured entirely by $B_0$, which reduces
to $(2\bar\sigma_+)^{-\Delta_-}$ up to a constant, consistent with
\eqref{eq:Kf0final}. At $n\geq 1$ both branches contribute on equal footing
and together account for the finite-$\mu$ corrections.

\section{$\Delta_+$ expansion of $K_f$}
\label{sec:IR correction}

This appendix derives the position-space IR expansion of $K_f$ in detail, starting from the momentum-space kernel \eqref{eq:K_f}.

In the IR regime $|k|^{2\nu}<\mu^{2\nu}$, the deformation factor admits the
convergent geometric expansion
\begin{equation}
\label{eq:IRgeometricseriesofk}
    \frac{|k|^\nu}{|k|^{2 \nu}+\mu^{2 \nu}}
    \,=\,
    \frac{|k|^\nu}{\mu^{2\nu}}\,\frac{1}{1+|k|^{2\nu}/\mu^{2\nu}}
    \,=\,
    \sum_{n=0}^{\infty}
    (-1)^n\,\mu^{-2(n+1)\nu}\,|k|^{(2n+1)\nu},
\end{equation}
organized in powers of $(|k|/\mu)^{2\nu}$, mirroring \eqref{eq:UVgeometricseriesofk}
with the roles of $|k|$ and $\mu$ exchanged.

Substituting into \eqref{eq:K_f} and Fourier-transforming term by term using the
standard radial identity \cite{GradshteynRyzhik}, one obtains
\begin{align}
\label{eq:IRkernel}
K_f(z,r)
\,=\,
\frac{1}{2\nu c_+}\sum_{n=0}^{\infty}
&(-1)^n\,\mu^{-2(n+1)\nu}\,\bar{\mathcal{N}}_n\,
\frac{\Gamma\!\left(d/2+n\nu\right)\Gamma\!\left(d/2+(n+1)\nu\right)}{\Gamma(d/2)}\notag\\
&\times\,z^{-d/2-(2n+1)\nu}\,
{}_2F_1\!\left(d/2+n\nu,\,d/2+(n+1)\nu;\,d/2;\,-\frac{r^2}{z^2}\right),
\end{align}
with the $n$-dependent prefactor
\begin{equation}\label{eq:barNn-def}
    \bar{\mathcal{N}}_n
    \,\equiv\,
    \frac{2^{(2n+1)\nu-1}}{\pi^{d/2}}\,.
\end{equation}
As in the UV appendix, the factor
$\Theta(\bar\sigma_+)$ is suppressed throughout.

For $n=0$, the hypergeometric function simplifies via
${}_2F_1(a,b;a;x)=(1-x)^{-b}$, giving
\begin{equation}
    {}_2F_1\!\left(d/2,\,d/2+\nu;\,d/2;\,-\frac{r^2}{z^2}\right)
    \,=\,
    \left(1+\frac{r^2}{z^2}\right)^{-\Delta_+}.
\end{equation}
Using $\Delta_\pm=d/2\pm\nu$ and $2\bar\sigma_+ = (z^2+r^2)/z$, the leading
term becomes
\begin{equation}\label{eq:Kf0IRfinal}
    K_f^{(0)}(z,r)
    \,=\,
    \frac{\mu^{-2\nu}}{2\nu c_+}\,
    \frac{2^{\nu-1}\,\Gamma(\Delta_+)}{\pi^{d/2}}\,
    (2\bar\sigma_+)^{-\Delta_+}\,,
\end{equation}
proportional to the undeformed standard-quantization bulk-to-boundary
propagator $K_{\Delta_+}$, suppressed by the explicit $\mu^{-2\nu}\propto
f^{-1}$ factor characteristic of the IR fixed point. The higher-order terms
are suppressed by $(|k|/\mu)^{2n\nu}$ in momentum space and encode the
deformation away from the IR fixed point.

Applying the ${}_2F_1$ connection formula \cite{DLMF} to
\eqref{eq:IRkernel} with
\begin{equation}
    a=d/2+n\nu,\qquad b=d/2+(n+1)\nu,\qquad c=d/2,
\end{equation}
and using
\begin{equation}
    1-\left(-\frac{r^2}{z^2}\right)\,=\,\frac{2\bar\sigma_+}{z}\,,
\end{equation}
the kernel splits into two branches:
\begin{equation}\label{eq:IRkernelTwoBranches}
K_f(z,\bar\sigma_+)
\,=\,
\frac{1}{2\nu c_+}\sum_{n=0}^{\infty}
(-1)^n\,\mu^{-2(n+1)\nu}\,\bar{\mathcal{N}}_n\,
\bigl[\bar A_n(z,\bar\sigma_+) + \bar B_n(z,\bar\sigma_+)\bigr],
\end{equation}
where
\begin{align}
\label{eq:IRAn-app}
\bar A_n(z,\bar\sigma_+)
\,=\,
&\,(2\bar\sigma_+)^{-d/2-(2n+1)\nu}\,\Gamma\!\left(d/2+(2n+1)\nu\right)\notag\\
&\times\,{}_2F_1\!\left(-n\nu,\,-(n+1)\nu;\,1-d/2-(2n+1)\nu;\,\frac{2\bar\sigma_+}{z}\right),\\
\label{eq:IRBn-app}
\bar B_n(z,\bar\sigma_+)
\,=\,
&\,z^{-d/2-(2n+1)\nu}\,
\frac{\Gamma\!\left(d/2+n\nu\right)\Gamma\!\left(d/2+(n+1)\nu\right)\Gamma\!\left(-d/2-(2n+1)\nu\right)}{\Gamma(-n\nu)\,\Gamma(-(n+1)\nu)}\notag\\
&\times\,{}_2F_1\!\left(d/2+n\nu,\,d/2+(n+1)\nu;\,1+d/2+(2n+1)\nu;\,\frac{2\bar\sigma_+}{z}\right).
\end{align}

Compared with the UV decomposition \eqref{eq:KfUVsigmabranches}, the roles of
the two branches are mechanically reversed: in the IR expansion it is
$\bar A_n$ that carries the boundary singularity $\bar\sigma_+^{-d/2-(2n+1)\nu}$,
while $\bar B_n$ depends on the finite bulk depth $z$. At $n=0$, the bulk-regular branch $\bar B_0$ vanishes by the same
reason that its prefactor contains
$1/[\Gamma(0)\Gamma(-\nu)]=0$. The leading IR contribution is therefore
captured entirely by $\bar A_0\propto (2\bar\sigma_+)^{-\Delta_+}$, consistent
with \eqref{eq:Kf0IRfinal}. At $n\geq 1$ both branches contribute on equal
footing and together account for the finite-$f$ corrections.

\section{Perturbative QNM shifts at the endpoints of the Robin flow}
\label{sec:app_perturbative-shifts}

The QNM trajectories $\tilde\omega(g)$ approach the alternate spectrum as
$g \to 0$ and the standard spectrum as $g \to \infty$. In both limits the
shift away from the endpoint is computable in closed form by expanding the
master equation
\begin{equation}
\label{eq:H-master-app}
    H(\tilde\omega;g,\tilde k,\nu)
    = \mathcal B(\tilde\omega,\tilde k;\nu)
    + g\,\mathcal A(\tilde\omega,\tilde k;\nu)
    = 0
\end{equation}
around the zeros of $\mathcal B$ and $\mathcal A$ respectively. We treat
the generic $\tilde k \neq 0$ case in Section \ref{subsec:app_k-nonzero}, where
both amplitudes have simple zeros and the QNM shift is linear in $g$ (or
in $1/g$); the special case $\tilde k = 0$ is treated in
Section \ref{subsec:app_k-zero}, where the zeros are double and the shift is of
square-root type.

\subsection{Generic momentum: linear perturbation}
\label{subsec:app_k-nonzero}

For $\tilde k \neq 0$, the amplitude $\mathcal B(\tilde\omega,\tilde k;\nu)$
has simple zeros at
\begin{equation}
\label{eq:alt-zeros}
    \tilde\omega^{(\mathrm{alt})}_{m,\pm}
    = \pm \tilde k - i(1 - \nu + 2m),
    \qquad m = 0, 1, 2, \ldots .
\end{equation}
These arise from the poles of the gamma functions in the denominator of
$\mathcal B$: $\tilde\omega = + \tilde k - i(1-\nu+2m)$ corresponds to the
argument $(1-\nu)/2 - i(\tilde\omega - \tilde k)/2 = -m$ becoming a
non-positive integer in the first gamma factor of \eqref{eq:AB}, and
similarly $\tilde\omega = -\tilde k - i(1-\nu+2m)$ from the second.

Let $\tilde\omega = \tilde\omega^{(\mathrm{alt})}_{m,+} + \delta\tilde\omega$
with $\delta\tilde\omega = O(g)$. To leading order in $g$,
\begin{equation}
\label{eq:B-expand-small-g}
    \mathcal B(\tilde\omega;\tilde k,\nu)
    = \partial_{\tilde\omega} \mathcal B
    \big|_{\tilde\omega^{(\mathrm{alt})}_{m,+}}\,\delta\tilde\omega
    + O(\delta\tilde\omega^2),
\end{equation}
while $\mathcal A$ is evaluated at the unshifted alternate frequency. The
master equation \eqref{eq:H-master-app} then gives
\begin{equation}
\label{eq:delta-w-linear}
    \delta\tilde\omega
    = -\frac{g\,\mathcal A(\tilde\omega^{(\mathrm{alt})}_{m,+};
        \tilde k,\nu)}
        {\partial_{\tilde\omega}\mathcal B(\tilde\omega^{(\mathrm{alt})}_{m,+};
        \tilde k,\nu)}
    + O(g^2).
\end{equation}

To evaluate the right-hand side, use the residue structure of $1/\mathcal B$
near the pole. Writing
\begin{equation}
    \mathcal B(\tilde\omega;\tilde k,\nu) =
    \frac{1}{\Gamma \left(\tfrac{1-\nu}{2} - \tfrac{i(\tilde\omega+\tilde k)}{2}\right)
        \Gamma \left(\tfrac{1-\nu}{2} - \tfrac{i(\tilde\omega-\tilde k)}{2}\right)},
\end{equation}
the zero at $\tilde\omega = +\tilde k - i(1-\nu+2m)$ comes from the second
gamma factor having argument $-m$. Using $\Gamma(z) \sim (-1)^m/(m!\,(z+m))$
near $z = -m$, one finds
\begin{equation}
    \partial_{\tilde\omega}\mathcal B
    \big|_{\tilde\omega^{(\mathrm{alt})}_{m,+}}
    = \frac{-i/2 \cdot (-1)^m \cdot m!}
        {\Gamma \left(\nu - m - i\tilde k\right)}.
\end{equation}
Meanwhile $\mathcal A$ evaluated at the alternate zero gives
\begin{equation}
    \mathcal A(\tilde\omega^{(\mathrm{alt})}_{m,+};\tilde k,\nu)
    = \frac{1}{\Gamma(\nu - m)\,\Gamma(\nu - m - i\tilde k)},
\end{equation}
using
$\tfrac{1+\nu}{2} - \tfrac{i(\tilde\omega+\tilde k)}{2} = \nu - m - i\tilde k$
and
$\tfrac{1+\nu}{2} - \tfrac{i(\tilde\omega-\tilde k)}{2} = \nu - m$
at this zero. Substituting into \eqref{eq:delta-w-linear},
\begin{equation}
\label{eq:seed-R-derived}
    \delta\tilde\omega_{m,+}
    = -\frac{2(-1)^m\,i\,g\,\Gamma(-m-i\tilde k)}
        {m!\,\Gamma(\nu - m)\,\Gamma(\nu - m - i\tilde k)}
    + O(g^2).
\end{equation}
This is the form adopted as the small-$g$ seed in the numerical continuation.
The left-mover shift $\delta\tilde\omega_{m,-}$ is obtained by sending
$\tilde k \to -\tilde k$ in \eqref{eq:seed-R-derived}.The shift is generically complex and order-$g$. Its real part displaces
the QNM along the chirality line $\text{Re}\,\tilde\omega = \pm\tilde k$,
and its imaginary part is the leading double-trace renormalization of the
QNM damping rate.

The same analysis at $g \to \infty$ is obtained by rewriting
$H = \mathcal B + g \mathcal A = g(\mathcal A + g^{-1}\mathcal B)$ and
expanding around the simple zeros of $\mathcal A$,
\begin{equation}
    \tilde\omega^{(D)}_{m,\pm}
    = \pm\tilde k - i(1 + \nu + 2m),
    \qquad m = 0, 1, 2, \ldots .
\end{equation}
Setting $\tilde\omega = \tilde\omega^{(D)}_{m,+} + \delta\tilde\omega$ with
$\delta\tilde\omega = O(1/g)$ and repeating the residue argument with
$\mathcal A \leftrightarrow \mathcal B$ and $\nu \leftrightarrow -\nu$
(equivalently, $\Delta_- \leftrightarrow \Delta_+$),
\begin{equation}
\label{eq:large-g-shift}
    \delta\tilde\omega
    = -\frac{1}{g}\,
    \frac{\mathcal B(\tilde\omega^{(D)}_{m,+};\tilde k,\nu)}
        {\partial_{\tilde\omega}\mathcal A(\tilde\omega^{(D)}_{m,+};\tilde k,\nu)}
    + O(1/g^2).
\end{equation}
Evaluating the gamma functions gives
\begin{equation}
\label{eq:standard-shift}
    \delta\tilde\omega_{m,+}^{\,(D)}
    = -\frac{2(-1)^m\,i\,\Gamma(-m - i\tilde k)}
        {g\,m!\,\Gamma(-\nu - m)\,\Gamma(-\nu - m - i\tilde k)}
    + O(1/g^2),
\end{equation}
which has the same structure as \eqref{eq:seed-R-derived} after the formal
substitution $\nu \to -\nu$ together with $g \to 1/g$. This relation reflects
the fact that the master equation $H = 0$ is invariant
under $(\mathcal A, \mathcal B) \leftrightarrow (\mathcal B, \mathcal A)$
together with $g \to 1/g$, and that $\mathcal A(\tilde\omega;\tilde k,\nu)
= \mathcal B(\tilde\omega;\tilde k,-\nu)$ as functions of $\nu$. The small-$g$
and large-$g$ perturbative regimes are therefore two faces of the same
calculation, expanded around the $\Delta_-$ and $\Delta_+$ poles of the
boundary two-point function respectively. This is the perturbative
counterpart of the Klebanov--Witten relation between alternate and standard
quantization: the alternate spectrum at $g = 0$ and the standard spectrum at
$g = \infty$ are related by $\Delta_- \leftrightarrow \Delta_+$, and the
Robin family is the one-parameter interpolation between them.

\subsection{Vanishing momentum: square-root perturbation}
\label{subsec:app_k-zero}

At $\tilde k = 0$ the two gamma factors in each of $\mathcal A$ and
$\mathcal B$ coincide. The zeros become double, and linear perturbation
theory breaks down. Writing the alternate amplitude near
$\tilde\omega_0 = -i(1-\nu+2m)$,
\begin{equation}
    \mathcal B(\tilde\omega;0,\nu)
    = \frac{1}{\Gamma \left(\tfrac{1-\nu}{2} - \tfrac{i\tilde\omega}{2}\right)^2},
\end{equation}
and using $\Gamma(z)^{-1} \sim (-1)^m m!\,(z+m)$ near $z = -m$,
\begin{equation}
    \mathcal B(\tilde\omega;0,\nu)
    = \tfrac{1}{4}\bigl(m!\bigr)^2\,(\tilde\omega - \tilde\omega_0)^2
    + O \bigl((\tilde\omega - \tilde\omega_0)^3\bigr).
\end{equation}
Meanwhile
\begin{equation}
    \mathcal A(\tilde\omega_0;0,\nu) = \frac{1}{\Gamma(\nu - m)^2}.
\end{equation}
The master equation $\mathcal B + g \mathcal A = 0$ at leading order
becomes
\begin{equation}
    \tfrac{1}{4}\bigl(m!\bigr)^2\,(\tilde\omega - \tilde\omega_0)^2
    + \frac{g}{\Gamma(\nu - m)^2}
    = 0,
\end{equation}
giving the $\sqrt{g}$ splitting
\begin{equation}
\label{eq:seed-k0-derived}
    \tilde\omega - \tilde\omega_0
    = \pm \frac{2\,i\sqrt{g}}{m!\,\Gamma(\nu - m)}
    + O(g).
\end{equation}
The leading factor of $i$ is the immediate origin of the closed-loop
trajectories observed in Fig. \ref{fig:qnm-trajectories}(a, b): for
small real $g > 0$ the displacement $\tilde\omega - \tilde\omega_0$ is
purely real, splitting the degenerate double-zero into two QNMs at
$\pm \text{Re}\,\tilde\omega \neq 0$ with the same imaginary part. The
two branches then evolve symmetrically through the right and left
half-planes and re-merge at the corresponding double zero of $\mathcal A$
as $g \to \infty$, where an analogous $\sqrt{1/g}$ splitting governs the
closing of the loop.

The structural reason for the square-root behavior at $\tilde k = 0$ is
that the alternate and standard QNMs are themselves order-2 EPs of the amplitudes $\mathcal B$ and $\mathcal A$ individually,
sitting at $g = 0$ and $g = \infty$ respectively. As $\tilde k$ is turned
on, the boundary degeneracy is resolved into two simple zeros, and the
square-root perturbation \eqref{eq:seed-k0-derived} smoothly becomes the
linear perturbation \eqref{eq:seed-R-derived}. The interior EP at finite
$g_c(\tilde k)$ studied in Section \ref{subsec:EP_transition} can be viewed
as the boundary EP at $\tilde k = 0$ migrating inward along the Robin
flow as $\tilde k$ and $\nu$ are tuned; the locus
$\bigl(\tilde k, \nu_c(\tilde k), g_c(\tilde k)\bigr)$ traces this
migration.

\subsection{Numerical seeds}
\label{subsec:app_seeds}

The closed-form shifts \eqref{eq:seed-R-derived}
and \eqref{eq:seed-k0-derived} provide the analytic starting points of
the numerical continuation. Setting $g = g_{\min} \sim 10^{-6}$ in each
formula gives a position infinitesimally displaced from the endpoint
QNM and accurate to $O(g_{\min}^2)$ or $O(g_{\min})$ respectively. The
Newton corrector then refines the seed to the exact zero of $H$ at
$g = g_{\min}$, and the trajectory is continued in logarithmically-spaced
$g$ steps to $g = g_{\max}$.

\acknowledgments
The authors gratefully acknowledge support from the Graduate School of Brown University. We thank David Lowe for valuable discussions.

\bibliographystyle{JHEP}
\bibliography{citation}
\end{document}